\documentclass[traditabstract]{aa} 
\usepackage{graphicx}

\usepackage{lscape}
\usepackage{amssymb}
\usepackage{amsmath}

\usepackage{txfonts}

\usepackage{booktabs}
\usepackage{subfigure}
\usepackage{placeins}
\usepackage{verbatim}
\usepackage{adjustbox}
\usepackage[toc,page]{appendix} 
\usepackage{tabularx}
\usepackage{lscape}
\usepackage{textcomp}
\usepackage{gensymb}
\usepackage{hyperref}
\usepackage{multicol}
\usepackage[symbol]{footmisc}
\usepackage{tikz}
\usepackage{caption}
\usepackage{longtable}

\usepackage{natbib}
\usepackage{supertabular}
\usepackage{xspace}
\usepackage{color}

\newcommand{\bd}{\begin{displaymath}}
\newcommand{\ed}{\end{displaymath}}
\newcommand{\be}{\begin{equation}}
\newcommand{\ee}{\end{equation}}
\newcommand{\beaa}{\begin{eqnarray*}}
\newcommand{\eeaa}{\end{eqnarray*}}
\newcommand{\bea}{\begin{eqnarray}}
\newcommand{\eea}{\end{eqnarray}}

\usepackage{xspace}

\def\macs1149{MACS 1149\xspace}
\def\HST{{\it HST\xspace}{}}

\def\GLEE{\textsc{Glee}\xspace}

\def\newAcand{24\xspace}
\def\newBcand{138\xspace}
\def\recovcand{118\xspace}

\defcitealias{canameras21b}{C21}
\defcitealias{rojas23}{R23}

\begin{document}

%
   \title{HOLISMOKES. XIII. Strong-lens candidates at all mass scales and their environments from the Hyper-Suprime Cam and deep learning}

  \titlerunning{Lens candidates of all mass scales}

   \author{S.~Schuldt\inst{1}\inst{,2}
            \and
            R.~Ca{\~n}ameras\inst{3}\inst{,4}\inst{,5}
            \and
            I.~T.~Andika\inst{4}\inst{,3}
            \and
            S.~Bag\inst{4}\inst{,3}
            \and
            A.~Melo\inst{3}\inst{,4}
            \and
            Y.~Shu\inst{6}
            \and
            S.~H.~Suyu\inst{4}\inst{,3}\inst{,7}
            \and\\
            S.~Taubenberger\inst{3}\inst{,4}
            \and
            C.~Grillo\inst{1}\inst{,2}
            }

        \institute{Dipartimento di Fisica, Universit\`a  degli Studi di Milano, via Celoria 16, I-20133 Milano, Italy\\
        e-mail: \href{mailto:stefan.schuldt@unimi.it}{\tt stefan.schuldt@unimi.it}
        \and
        INAF - IASF Milano, via A. Corti 12, I-20133 Milano, Italy
        \and
        Max-Planck-Institut f{\"u}r Astrophysik, Karl-Schwarzschild Stra{\ss}e 1, 85748 Garching, Germany
        \and
        Technical University of Munich, TUM School of Natural Sciences, Physics Department,  James-Franck-Stra{\ss}e 1, 85748 Garching, Germany
        \and
        Aix Marseille Univ, CNRS, CNES, LAM, Marseille, France
        \and
        Purple Mountain Observatory, No. 10 Yuanhua Road, Nanjing, Jiangsu, 210033, People’s Republic of China
        \and
        Academia Sinica Institute of Astronomy and Astrophysics (ASIAA), 11F of ASMAB, No.1, Section 4, Roosevelt Road, Taipei 10617, Taiwan
         }

   \date{Received --; accepted --}

 
  \abstract{We performed a systematic search for strong gravitational lenses using Hyper Suprime-Cam (HSC) imaging data, focusing on galaxy-scale lenses combined with an environment analysis resulting in the identification of lensing clusters. To identify these lens candidates, we exploited our residual neural network from HOLISMOKES~VI, trained on realistic $gri$ mock-images as positive examples, and real HSC images as negative examples. Compared to our previous work, where we successfully applied the classifier to around 62.5 million galaxies having an $i$-Kron radius of $\geq 0.8\arcsec$, we now lowered the $i$-Kron radius limit to $\geq 0.5\arcsec$. The result in an increase by around 73 million sources, amounting to a total of over 135 million images.
  During our visual multi-stage grading of the network candidates, we also simultaneously inspected larger stamps ($80\arcsec \times 80\arcsec$) to identify large, extended arcs cropped in the $10\arcsec \times 10\arcsec$ cutouts and also classify their overall environment. Here, we also re-inspected our previous lens candidates with $i$-Kron radii of $\geq 0.8\arcsec$ and classified their environment. Using the 546 visually identified lens candidates, we further defined various criteria by exploiting extensive and complementary photometric redshift catalogs to select the candidates in overdensities.
  In total, we identified \newAcand grade A and \newBcand grade B candidates that exhibit either spatially-resolved multiple images or extended, distorted arcs in the new sample. Furthermore, combining our different techniques to determine overdensities, we identified a total of 231/546 lens candidates by at least one of our three identification methods for overdensities. This new sample contains only 49 group- or cluster-scale re-discoveries, while 43 systems had been identified by all three procedures. Furthermore, we performed a statistical analysis by using the neural network from HOLISMOKES~IX to model these systems as singular isothermal ellipsoids with external shear and to estimate their parameter values,
  making this the largest uniformly modeled sample to date. We find a tendency towards larger Einstein radii for galaxy-scale systems in overdense environments, while the other parameter values as well as the uncertainty distributions are consistent between those in overdense and non-overdense environments.
  These results demonstrate the feasibility of downloading and applying neural network classifiers to hundreds of million cutouts, which will be needed in the upcoming era of big data from deep, wide-field imaging surveys such as Euclid and the \textit{Rubin} Observatory Legacy Survey of Space and Time. At the same time, it offers a sample size that can be visually inspected by humans. These deep learning pipelines, with false-positive rates of $\sim0.01$\%, are very powerful tools to identify such rare galaxy-scale strong lensing systems, while also aiding in the discovery of new strong lensing clusters.}


   \maketitle

\section{Introduction}
\label{sec:intro}

Strong gravitational lensing systems, both on galaxy and cluster scales, are very powerful tools used for probing the Universe's properties in various fields, such as the study of the nature and distribution of dark matter \citep[e.g.,][]{schuldt19, shajib21, wang22} and high-redshift systems \citep[e.g.,][]{lemon18, shu18, vanzella21, mestric22}. In particular, galaxy-scale lensing systems in clusters are ideal for studying the galaxy mass distribution in the presence of a large-scale halo, as well as to constrain the subhalo density profiles or the inner mass structure of galaxies at the low-end of the stellar mass and velocity dispersion functions \citep[e.g.,][]{grillo08, grillo14, parry16, granata23, despali24}. Furthermore, strong-lensing systems with time-variable objects, such as a quasar or supernova (SN), enable the measurement of the Hubble constant, $H_0$, and the geometry of the Universe as proposed by \citet{refsdal64}. So far, mostly galaxy-scale systems with lensed quasars have been exploited for this \citep[e.g., by the H0LiCOW and TDCOSMO collaborations, see e.g.,][]{wong20, birrer20, shajib22}; however, this was also demonstrated on cluster scales \citep[e.g.,][]{acebron22a, acebron22b}, where the mass distribution is significantly more complicated. Extending this systematically to lensed SNe is one of the main scientific goals of our Highly Optimized Lensing Investigations of Supernovae, Microlensing Objects, and Kinematics of Ellipticals and Spirals \citep[HOLISMOKES][]{suyu20} program. To date, only very few lensed SNe have been found and, in fact, only two systems, SN Refsdal and SN H0pe \citep{kelly15_ATel, kelly15, frye24}, lensed by galaxy clusters have sufficiently precise time delays to enable measurements of the Hubble constant and the geometry of the universe \citep{grillo18, grillo20, kelly23a, grillo24, pascale24}. A third lensed supernova is called SN Encore \citep{pierel24}, whose galaxy previously hosted SN Requiem \citep{rodney21}. It was discovered in November 2023 and the time-delay cosmography analyses with this system are ongoing (Ertl et al. in prep., Suyu et al. in prep).

With recently started and upcoming wide-field imaging surveys such as the \textit{Rubin} Observatory Legacy Survey of Space and Time \citep[LSST;][]{ivezic08} from the ground, complemented by the {\it Euclid} \citep{laureijs11} and {\it Roman} \citep{green12} satellites from space, the amount of astronomical imaging data sets is expected to increase significantly in the optical and near-infrared wavelength range over the next few years. Thanks to their runtime and performance in image pattern recognition, supervised deep learning (DL) techniques such as convolutional neural networks \citep[CNNs][]{lecun98} are playing a significant role in the analysis of these data sets. Once trained, these networks can be applied to millions or billions of image cutouts within an acceptable amount of time. Beside exploiting supervised CNNs for photometric redshift estimation \citep{DIsario18, schuldt21a, johnwilliam23, jones24}, modeling strong galaxy-scale lenses \citep[e.g.,][]{hezaveh17, pearson19, pearson21, schuldt21b, schuldt23a, schuldt23b}, detecting dark matter substructure \citep{tsang24}, estimating structural parameters of galaxies \citep{tuccillo18, tohill21, li22a}, or classifying them according to their morphology \citep{dieleman15, walmsley22}, DL became the state of the art technique for lens classification \citep{metcalf19}. Although they have shared the same baseline of relying on DL, there are various projects that have targeted different image sets \citep[e.g.,][]{jacobs17, jacobs19, canameras20, canameras23, savary22} and are possibly limited to particular lens samples, such as lensed quasars \citep[e.g.,][]{andika23}, systems with high lens redshift \citep[e.g.,][]{shu22}, or those situated within a cluster environment \citep{angora23}. These CNNs are complementing previous non-DL algorithms \citep[e.g.,][]{chan15, sonnenfeld18a, shu16a} with an overall better classification performance \citep{metcalf19}.

In \citet[hereafter \citetalias{canameras21b}]{canameras21b}, we presented a CNN trained on realistic mock images and applied it to $gri$ multiband images from the Hyper-Suprime Cam Subaru Strategic Program \citep[HSC-SSP;][]{aihara18a}, complemented with a detailed performance test presented by \citet{canameras23}. Avoiding any strict cuts on the catalog level, we applied it to 62.5 million image stamps from the HSC Wide layer corresponding to a lower limit on the $i$-band Kron radius of $0.8\arcsec$. This is possible thanks to a false positive rate (FPR) of $\sim 0.01$\%. From the 9 651 resulting systems, corresponding to 0.015\% of the input catalog, we applied our visual inspection to identify 88 secure (grade A) and 379 probable (grade B) lens candidates. Given the high success rate, in this work we offer new lens candidates by lowering the limit on the $i$-Kron radius to 0.5\arcsec, which gives us an additional $\sim$ 73 million image stamps.

Since the scope of \citetalias{canameras21b} was specifically set on the detection of new galaxy-scale lenses, we neglected, as is commonly done, possible group- or cluster-scale lensing features during our visual inspection. However, as previously reported in \citetalias{canameras21b}, our network was able to recover also group- and cluster-scale lenses. Therefore, we adjusted our visual inspection strategy and this time we also specifically report lenses in a significantly overdense environment, such as a galaxy cluster, by inspecting larger cutouts ($80\arcsec$ on a side). This was complemented by a re-inspection of the lens candidates from \citetalias{canameras21b} to classify the environment. 

While the deep learning-based identification of galaxy-scale lenses in the field has been well explored, only \citet{angora23} has developed a network to identify galaxy-scale lenses in galaxy clusters, which was trained on $4\arcsec \times 4\arcsec$ high-resolution image cutouts in known clusters. In contrast, we used $\sim 12\arcsec \times 12\arcsec $ ground-based image stamps from HSC, enabling the network to analyze the close environment and identify extended arcs influenced by a galaxy cluster. Furthermore, we applied it to any astronomical source targeted by HSC with $i$-Kron radius above 0.5\arcsec, which enabled the identification of new galaxy-scale lenses in the field as well. 

Despite the scarcity of CNN-based searches for galaxy-scale systems in clusters, many known lensing clusters include also galaxy-scale systems. In fact, as shown by \citet{meneghetti20, meneghetti22, meneghetti23}, around one order of magnitude more galaxy-scale systems in clusters have been observed than are expected based on hydrodynamic simulations in a $\Lambda$CDM cosmology.

Using the visually identified lens candidates in overdensities and lens candidates near known galaxy clusters, we have been able to define and test several selection criteria for their identification. These criteria are based on the photometric redshift distribution of their surrounding objects. Here, we focus specifically on our lens candidates, as lenses are already on average in slightly overdense environments \citep[e.g.,][]{wells24}. We exploited three complementing photometric redshift catalogs, providing more than a hundred million measurements in the targeted footprint. Thanks to the expected accurate and large photometric redshift catalog from ongoing and upcoming wide-field imaging surveys, these criteria can be used to identify lenses in significant overdensities, while also identifying galaxy clusters independent of their lensing nature.

Going beyond the identification of galaxy-scale systems, while separating them into systems in an overdensity from those in the field, we applied the neural network presented by \citet{schuldt23a} to all our grade A and B candidates. We show the distributions for all seven predicted parameters of the adopted singular isothermal ellipsoid (SIE) plus external shear profiles. We then discuss the differences between systems in an overdensity and in the field.

The paper is organized as follows. We first introduce the overall procedure in Sect.~\ref{sec:methodology}. Sect.~\ref{sec:inspection} describes our visual inspection strategy and the identified lens candidates. In Sect.~\ref{sec:environment}, we highlight our analysis of the lens environment and show a  statistical analysis of their mass model parameters in Sect.~\ref{sec:massmodel}. Finally, we conclude our findings in Sect.~\ref{sec:conclusion}. Following \citetalias{canameras21b}, we have adopted the flat concordant $\Lambda$CDM cosmology with $\Omega_\text{M}$ $ = 1 - \Omega_\Lambda$ $ = 0.308$ \citep{planck16}, and with $H_0 = 72\, \text{km\,s}^{-1}\, \text{Mpc}^{-1}$ \citep{bonvin17}.

\FloatBarrier
\section{Methodology}
\label{sec:methodology}

We made use of the residual neural network presented by \citetalias{canameras21b}, and we refer to that publication as well as to \citet{canameras23}, for further details. In the following, we give a short summary of the network architecture, the ground truth data set, and the training procedure.

\subsection{Network architecture}
\label{sec:methodology:network}

Over time, neural networks have gotten more and more powerful thanks to the increasing extend of training sets, more powerful computing techniques that allows us to train deeper networks, improving their broad applicability, and other aspects. Consequently, different network architectures and algorithms were developed to optimize the performance. While the architecture for image processing tasks are overall still following the original setup of a simple CNN \citep{lecun98}, which consists of multiple convolutional layers followed by a number of fully connected layers, the depth of the network got significantly increased over time. In particular, the residual neural network (ResNet) concept \citep{he16a} introduces so-called skip-connections \citep[preactivated bottleneck residual units in][]{he16b} to allow for deep CNNs, avoiding vanishing gradient, so that we can aptly optimize the first layers, while keeping the computational costs at an acceptable level. Such ResNets have obtained excellent results on the ImageNet Large Scale Visual Recognition Challenge 2015 \citep{he16a}. In the recent past, these ResNets were also used for lens finding \citep[e.g.,][]{lanusse18, li20, huang21} and outperformed classical CNNs from the lens finding challenge \cite{metcalf19}. 

We used a ResNet whose architecture setup is based on the ResNet-18 architecture \citep{he16a}. It is composed of eight ResNet blocks with each two convolutional layers, batch normalization, and a ReLu activation function. These layers are followed by a fully connected layer with 16 neurons, before the last layer with a single neuron that outputs a score $p$ in the range [0,1] through to a sigmoid activation function.

\subsection{Ground truth dataset}
\label{sec:methodology:goundtruth}

Our binary classification network is trained and validated on a set of images composed by 40 000 positive and 40 000 negative examples. As positive examples, we used simulated images of galaxy-scale lenses that are based on real HSC $gri$ images of luminous red galaxies (LRGs). Following the procedure described in \citet{canameras20} and \citet{schuldt21a}, we added to those LRG images some arcs simulated with \GLEE \citep{suyu10a, suyu12b} from galaxy images of the Hubble Ultra Deep Field \citep{inami17} as background sources. Here, we adopted a SIE profile and estimated the ellipticity based on that of the light distribution. The Einstein radius was inferred from the corresponding velocity dispersion and redshift measurement taken from the Sloan Digital Sky Survey program \citep[SDSS,][]{SDSS14}. Since it is important to have a uniform distribution in the training set, we increased the fraction of systems with wide image separations to obtain a sample with a uniform Einstein radius distribution in the range of 0.75\arcsec to 2.5\arcsec. We also used a similar fraction of quadruply and doubly imaged systems.

This data set is complemented with negative examples, containing mostly spirals from \citet{tadaki20}, isolated LRGs, and random galaxies with similar proportions. These were selected from random sky positions of the HSC Wide footprint to mitigate impact from small-scale and depth variations.

\subsection{Training procedure}
\label{sec:methodology:training}

The data set described in Sect.~\ref{sec:methodology:goundtruth} was split up into 80\% training and 20\% validation sets. The test sample and performance tests are summarized in Sect.~\ref{sec:methodology:performance}. 

After a random initialization, the network was trained over 100 epochs, while the network was saved at the epoch with the minimal binary-entropy loss on the validation set. We used mini-batch stochastic gradient descent with 128 images per batch, a learning rate of 0.0006, a weight decay of 0.001, and a momentum of 0.9. In each epoch all images were randomly shifted by up to $\pm5$ pixels in $x$ and $y$ direction to improve the generalization. 

\subsection{Performance tests}
\label{sec:methodology:performance}

We carried out several performance test of the network based on a test set with HSC Wide PDR2 images as described in detail by \citet{canameras23}. In short, the completeness was tested with grade A or B galaxy-scale lenses from the Survey of Gravitationally-lensed Objects in HSC Imaging \citep[SuGOHI,][hereafter, SuGOHI sample]{sonnenfeld18a, sonnenfeld19, sonnenfeld20, wong20, chan20, jaelani20a, jaelani21a}. We excluded systems with high image separations (Einstein radius $\geq 4\arcsec$) from this sample, which did not match our training data and would consequently be likely to be missed. 

On the other hand, the FPR was estimated with a set of non-lens galaxies from the COSMOS field \citet{scoville07} by ensuring the exclusion of all known lenses and candidate lenses from the MasterLens database\footnote{\url{http://admin.masterlens.org}}, \citet{faure11}, \citet{pourrahmani18}, as well as systems from the SuGOHI sample. As shown in \citetalias{canameras21b}, using a threshold of $p\geq 0.1$, we obtained a completeness above 50\% on galaxy-scale systems and a FPR $\leq 0.01\%$. The obtained receiver operating characteristic curve is shown in Fig.~2 of \citetalias{canameras21b}. Examples of missed lenses are shown in Fig. 3 of \citetalias{canameras21b} and Fig. B.2 of \citet{canameras23}. These are typically lensing systems with atypical colors for the lens or source (e.g., red arcs), highly blended lenses, or lenses with significant contamination from other objects near the sightline.

Given our focus this time on lenses at all mass scales, approximately 25\% of the grade A and B SuGOHI group and cluster-scale lenses are recovered in our list of candidates. 
This relatively low recovery rate is understandable given that the network is trained specifically on galaxy-scale systems, but highlights the possibility of new group- and cluster lens discoveries with this ResNet, as well as generally with automated algorithms.

\FloatBarrier
\section{Lens grading}
\label{sec:inspection}

The following section describes the visual inspection and grading procedure of the network candidates with $i$-Kron radii between $0.5\arcsec$ and $0.8\arcsec$. We further discuss the outcome and the effect of the individual steps, which will help in planning future visual inspection strategies. We offer a comparison between \citetalias{canameras21b} and this work of the resulting numbers of network candidates, grade A, and B lens candidates in Table~\ref{tab:statisticsoverview}. Since we used the same network, the differences are coming from the different cut on the $i$-Kron radius when defining the parent sample.

\begin{table}[t!]
        \caption{Sample sizes at various stages of the analysis, including rediscoveries.\label{tab:statisticsoverview}}
    \begin{center}
    \begin{tabular}{c|cc}
   Description & \citetalias{canameras21b} & This work \\
   \noalign{\smallskip}\hline \hline \noalign{\smallskip}
    analysed sources (in million) & $\sim 62.5$ & $\sim 73$ \\ 
    network candidates ($p\geq0.1$) & 9 651 & 11 816\\ 
    network candidates after first cleaning & 2831 & 1 475 \\
    grade A lenses & 88 & 24 \\
    grade B lenses & 379 & 138 \\ \hline
    grade A or B & \multicolumn{2}{c}{546} \\ 
    grade A or B \& OD$_\text{vis}$ & \multicolumn{2}{c}{84} \\
    grade A or B \& OD$_\text{lit}$ & \multicolumn{2}{c}{174} \\
    grade A or B \& OD$_\text{phot}$ & \multicolumn{2}{c}{136} \\
    grade A or B \& any OD flag & \multicolumn{2}{c}{231} \\
    grade A or B \& all OD flags & \multicolumn{2}{c}{43}\\  \hline
   \end{tabular}
   \end{center}
   \textbf{Note.} The comparison between \citetalias{canameras21b}, analyzing sources with $i$-Kron radius above $0.8\arcsec$, to this work, with sources having an $i$-Kron radius between $0.5\arcsec$ and $0.8\arcsec$, clearly shows that larger sources are more likely acting as strong lenses, as expected. We note that this is a fair comparison as we use the same network and threshold, apart from the number of network candidates after first cleaning because of a less restrictive cleaning in \citetalias{canameras21b}. Further details and the exact definition of our overdensity (OD) flags, namely, OD$_\text{lit}$ based on the literature, OD$_\text{vis}$ based on the visual inspection, and OD$_\text{phot}$ on the photomtric redshifts, are given in this section and in Sect.~\ref{sec:environment}. The difference between the sum of grade A and B lens candidates from \citetalias{canameras21b} and this work (totaling to 629 systems) and the number of 546 systems analyzed in Sect.~\ref{sec:environment}, is the result of excluding duplicates.
\end{table}

\subsection{Network candidates}
\label{sec:inspection:network}

The trained network was applied to all $\sim$72 million images from the HSC Wide layer showing objects with an $i$-band Kron radius between 0.5\arcsec and 0.8\arcsec, complementing the sample from \citetalias{canameras21b} targeting objects with $i$-band Kron radii $\geq 0.8\arcsec$. Based on the tests in \citetalias{canameras21b}, we considered all 11~816 objects ($\sim $ 0.016\% of the input catalog) with a network score $ p\geq 0.1$ as network candidates. The obtained fraction (see Table~\ref{tab:statisticsoverview}) is comparable to that achieved in \citetalias{canameras21b}, which indicates that the network has the capability to also handle objects that appear smaller or with some offset to the image center, as several times the central object is one of the possible lensed images or a compact object near the candidate lens system.

\subsection{First cleaning through visual inspection}
\label{sec:inspection:firstcleaning}

As in previous lens search studies, including \citetalias{canameras21b}, these network candidates have a significant fraction of false positives. Thus, we decided to carry out a visual inspection of these systems using pre-generated $gri$-color images. These images were inspected first by one person to exclude obvious non-lenses, reducing the candidates to 1 475 systems. These false positives had mostly a low network score, such that a higher threshold would reduce the FPR significantly. However, in this case we would have missed several possible lens candidates. In fact, a threshold of 0.11 (instead of 0.1) would already exclude 12 grade B candidates, of which 8 are already known (see Table~\ref{tab:newcand} and Sect.~\ref{sec:inspection:finalcompilation}), justifying the relatively low threshold.

\begin{table*}[t!]
    \caption{High-confidence lens candidates with ResNet scores $p\geq 0.1$, and average grades $G\geq 1.5$ from visual inspection.}
    \begin{center}
    \begin{tabular}{c|cc|cccc|ccc|ccc|c}
Name   &RA [deg]  &Dec [deg]  & $p$ & $G$ & $\sigma_\text{G}$ & $z$ & OD$_\text{lit}$ & OD$_\text{vis}$ & OD$_\text{z}$ & $N_\text{max}$ & $z_\text{low}$ & $N_\text{tot}$ & References\\
(1)& (2) & (3) & (4) & (5) & (6) & (7) & (8) & (9) & (10) & (11) & (12) & (13) & (14) \\ \hline \hline \noalign{\smallskip}
 HSCJ2331$+$0037 & $ 352.87702 $ & $  0.62594 $ & $ 0.91 $ & $ 3.00 $ & $ 0.00 $ & 0.56 & Y & N & N & 18 & 0.56 & 724 & W18 C21\\
 HSCJ2305$-$0002 & $ 346.34029 $ & $ -0.03658 $ & $ 1.00 $ & $ 3.00 $ & $ 0.00 $ & 0.49 & N & N & Y & 22 & 0.62 & 609 & W18 C21 \\
 HSCJ0128$+$0038 & $  22.47583 $ & $  0.63363 $ & $ 0.14 $ & $ 2.88 $ & $ 0.33 $ & 0.61 & Y & N & Y & 25 & 0.78 & 546 & C21 S22 \\
 HSCJ0925$+$0017 & $ 141.43748 $ & $  0.28450 $ & $ 0.20 $ & $ 2.88 $ & $ 0.33 $ & 1.36 & Y & N & Y & 24 & 0.76 & 833 & J20 C21 \\
 HSCJ2332$+$0038 & $ 353.12887 $ & $  0.63939 $ & $ 0.98 $ & $ 2.75 $ & $ 0.43 $ & 0.61 & N & N & N & 20 & 0.80 & 779 & W18 C21\\
 \vdots & \vdots & \vdots & \vdots & \vdots &  \vdots &  \vdots &  \vdots &  \vdots &  \vdots &  \vdots &  \vdots &  \vdots &  \vdots\\ \hline
    \end{tabular}
    \end{center}
\textbf{Note.} Column description: (1) Source name, (2) right ascension (J2000), (3) declination (J2000), (4) network score, (5) average grade from our visual inspection, (6) dispersion among the eight graders, (7) photometric redshift of the recentered lens galaxy candidate from our combined catalog based on DEmP \citep{hsieh14}, Mizuki \citep{tanaka18}, and NetZ \citep{schuldt21b}, flag on the overdensity (OD) based on (8) Literature \citep{oguri14, oguri18, wen18, wen21}, (9) visual inspection, and (10) the photometric redshift analysis, (11) absolute height of photo-$z$ distribution, (12) photo-$z$ lower bound, (13) total number of photo-$z$ in considered area, and (14) references of their previous discovery with 
M16 for \citet{more16b},
D17 for \citet{diehl17},
S18 for \citet{sonnenfeld18a}, 
W18 for \citet{wong18}, 
H19 for \citet{huang19},
P19 for \citet{petrillo19b},
Ca20 for \citet{cao20}
H20 for \citet{huang20}
J20 for \citet{jaelani20b}, 
S20 for \citet{sonnenfeld20}, 
C21 for \citet{canameras21b},
R22 for \citet{rojas22}, 
S22 for \citet{shu22},
A23 for \citet{andika23},
J23 for \citet{jaelani23}, 
and ML for the master lens catalog \url{http://admin.masterlens.org}.
\\
    \label{tab:newcand}
\end{table*}

\subsection{Multiple grader inspection}
\label{sec:inspection:multiplegrader}

According to \citet[][hereafter \citetalias{rojas23}]{rojas23}, the visual inspection should be done by at least six individual persons to ensure stable average grades, while the accuracy is still dropping with increasing number of graders (compare Fig.~15 in \citetalias{rojas23}). To reduce the images that require inspection, we carried out a multi-step inspection with eight individuals\footnote{The visual inspectors are in alphabetical order by last name: I.~T.~A., S.~B., R.~C., A.~M., S.~S., Y.~S., S.~H.~S., and S.~T.}. 

During the visual grading, we inspected $gri$-color images with a size of $10\arcsec \times 10\arcsec$ to grade the central object and its close surrounding, showing possible arcs. In addition, which was done for the first time, we looked at larger images to classify their environment. Since 80\arcsec corresponds to about 0.5 Mpc at $z=0.5$, a typical lens cluster redshift \citep[e.g.,][]{bergamini21, acebron22a, acebron22b, schuldt24}, this size ensures that extended cluster-scale arcs remain within the images and enough of the cluster environment is included, while keeping the astronomical objects visible on the inspector's screen. Therefore, our grading tool shows, in addition to three $10\arcsec\times10\arcsec$ $gri$-color images with different filter scalings, also three $80\arcsec \times 80\arcsec$ cutouts centered at the small image stamp, again with different stretching factors. 

Since CNNs are translation invariant and we specifically include off-centered systems in our training set, the network also identifies strong-lensing systems offset from the center as long as the lensing features remain within the cutout. Thus it is not guaranteed that the lens is the central object. Also, sometimes the reported coordinate is not precise as the central object is blended with a neighboring object. Since we want to report the lens center, we also incorporated in our grading tool an option to indicate if the lens galaxy is not the central object. This is particularly important in our case, as multiple lens candidates are centered at their possible arc or at a neighbouring objects given our selection criteria on more compact objects. We corrected these offsets manually after grading for the most probable lenses reported in Table~\ref{tab:newcand}.

\textbf{Round \#1:} As initial step, we conducted a calibration round containing 200 systems inspected by everyone. We assigned to each system, as done in previous studies, an integer grade between zero and three, corresponding to ``not a lens'', ``possible lens'', `` probable lens'', and `` definite lens''. We then compared the 200 assigned grades from all eight individuals and discussed the systems with high discrepancies to understand the reasonings for giving different grades. This is crucial to calibrate the expectations and discuss individual systems to allow everyone to get familiar with the grading criteria and tool, as well as the specific images inspected (e.g., image resolution). Since lens candidates are sometimes ambiguous, especially for novice graders, even with a bullet-point list of criteria for the different grades, such discussions on the calibration set are helpful. The grades from round \#1 are not taken into account for the final lens candidate compilation shown in Table~\ref{tab:newcand}.

\begin{figure}
  \includegraphics[trim=0 0 0 0, clip, width=0.49\textwidth]{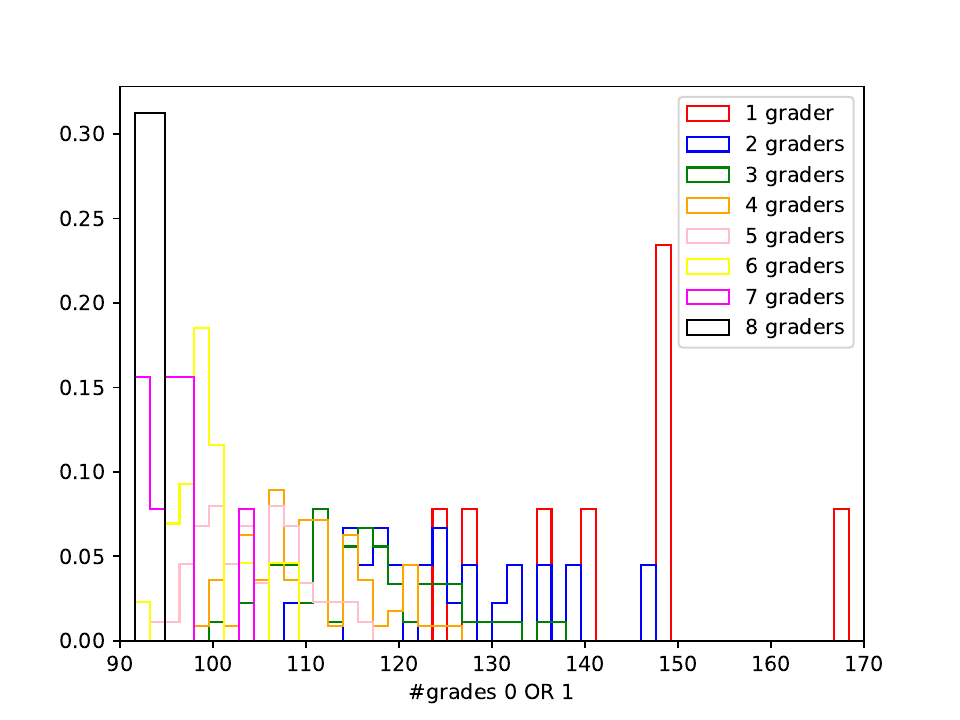}
  \caption{Normalized histograms showing the distribution of 0 and 1 grades assigned in round \#2 to network candidates from the overlap sample, containing 175 systems. We show all possible team combinations of the eight graders and only count systems that received a 0 or 1 by all graders of a particular team combination. Significant scatter is observed for small teams, which is expected and in agreement with \citetalias{rojas23}.}
  \label{fig:hist_grades01}
\end{figure}

\textbf{Round \#2:} After calibration, we split ourselves into two teams of each four people. Each team graded 825 of the 1 475 network candidates, with an overlap of 175 random systems (hereafter, the overlap sample) that are graded by both teams for comparison. As expected from \citetalias{rojas23}, we observe significant differences between the individual grades, but also between the two teams. In Fig.~\ref{fig:hist_grades01}, we show a normalized histogram of the number of grades 0 (``not a lens'') and grades 1 (``possible lens''), for all possible team combinations of the graders. We clearly see significant differences between a single grader and a team of two people. Significant scatter remains for three and four graders, which is in agreement with results from \citetalias{rojas23}, recommending to average over at least six individual grades.

\begin{figure}
  \includegraphics[trim=0 0 0 0, clip, width=0.5\textwidth]{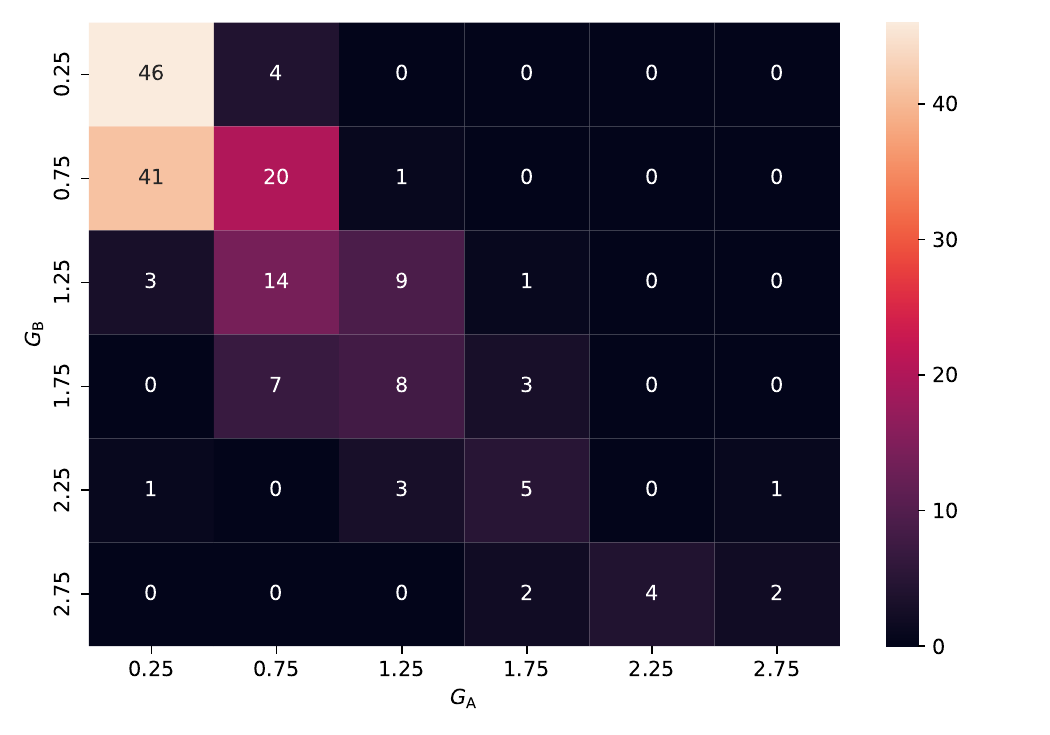}
  \includegraphics[trim=0 0 0 0, clip, width=0.5\textwidth]{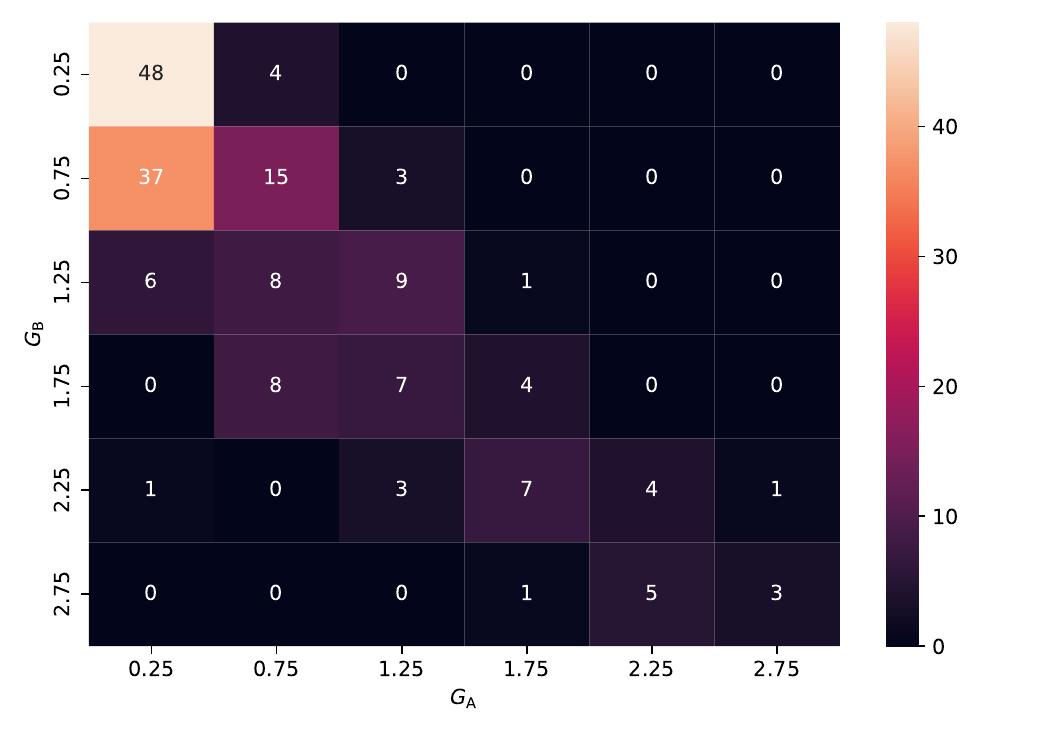}
  \includegraphics[trim=0 0 0 0, clip, width=0.5\textwidth]{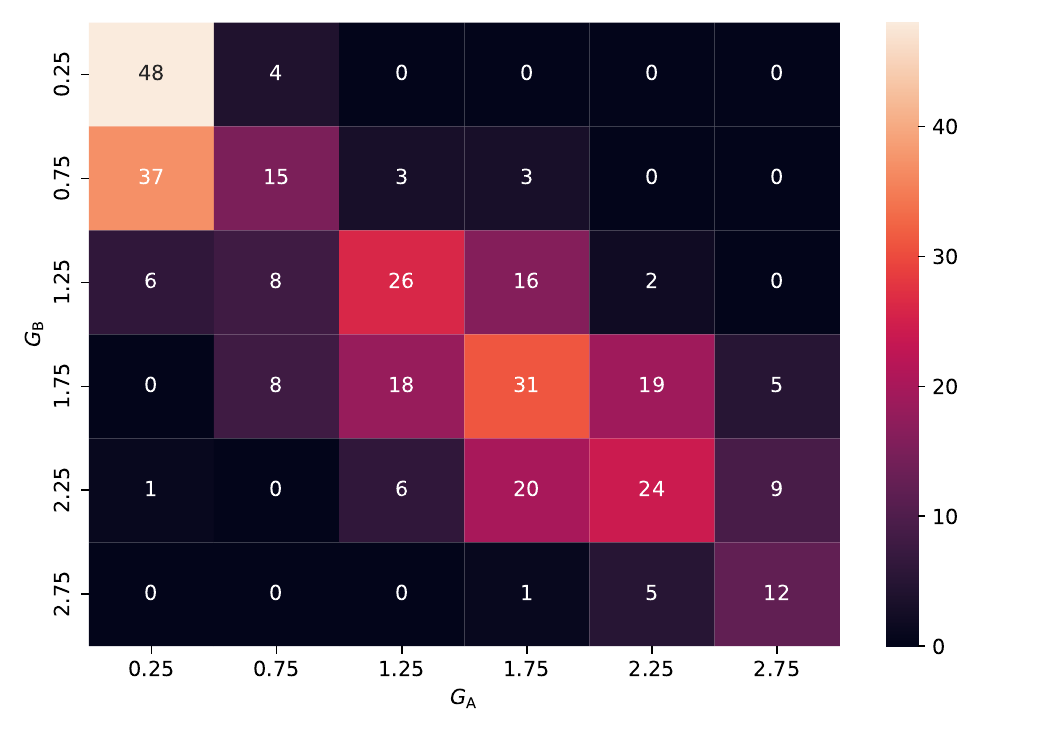}
  \caption{Comparison between average grades from team A ($x$-axis) and team B ($y$-axis), containing four graders apiece, from round \#2 (top), round \#3 (middle), each with 175 systems (overlap sample), and round \#4 (bottom), containing all 327 objects that received eight grades.}
  \label{fig:heatmaps}
\end{figure}

As mentioned above, we also observed differences between the two teams. This is shown in Fig.~\ref{fig:heatmaps}, where we plot the average grade $G_\text{A}$ from team A on the $x$-axis and the average grade $G_\text{B}$ from team B on the $y$-axis. Although each team contains four individuals, we see a tendency towards lower grades from team A. However, we note that the sample only contains 175 systems and thus, particularly at higher grades, the statistic is relatively poor. In detail, the top panel of Fig.~\ref{fig:heatmaps} shows the results from round \#2, while the middle panel displays the results from round \#3, and the bottom panel from round \#4, which we describe below.

\textbf{Round \#3}: Both teams re-inspected the systems with strong discrepancy in the provided grades within each team (standard deviation above 0.75). Cases with high dispersion in visual grades often show ambiguous blue arcs that could either be lensed arcs from background galaxies (without clear counter-images), spiral arms, or tidal features. From Fig.~\ref{fig:heatmaps}, we see a slight improvement from round \#2 to \#3 in terms of agreement in the average grades from the two teams, although the discrepancy remains.

\textbf{Round \#4}: To increase the number of examiners per object as recommended by \citetalias{rojas23}, each team graded the systems with average grade above 1 obtained by the complementary team. This excludes a significant fraction of additional false-positive candidates and systems with not very clear lensing features. This consequently reduces the time required for round \#4, while ensuring a stable average grade from eight inspectors per relevant object. We explicitly chose a slightly lower threshold than for the final compilation (see Sect.~\ref{sec:inspection:finalcompilation}) to counterbalance the observed discrepancy and the effect of lower number of inspectors.

In the bottom panel of Fig.~\ref{fig:heatmaps}, we show the differences between the final average grades for all 327 systems that obtained eight grades, again split into the previous two teams for comparison. This highlights the good agreement between both teams for the relevant systems and shows a stabilization for more than four graders.

\subsection{Final lens candidate compilation}
\label{sec:inspection:finalcompilation}

\begin{figure*}[t!]
  \includegraphics[trim=0 0 0 0, clip, width=\textwidth]{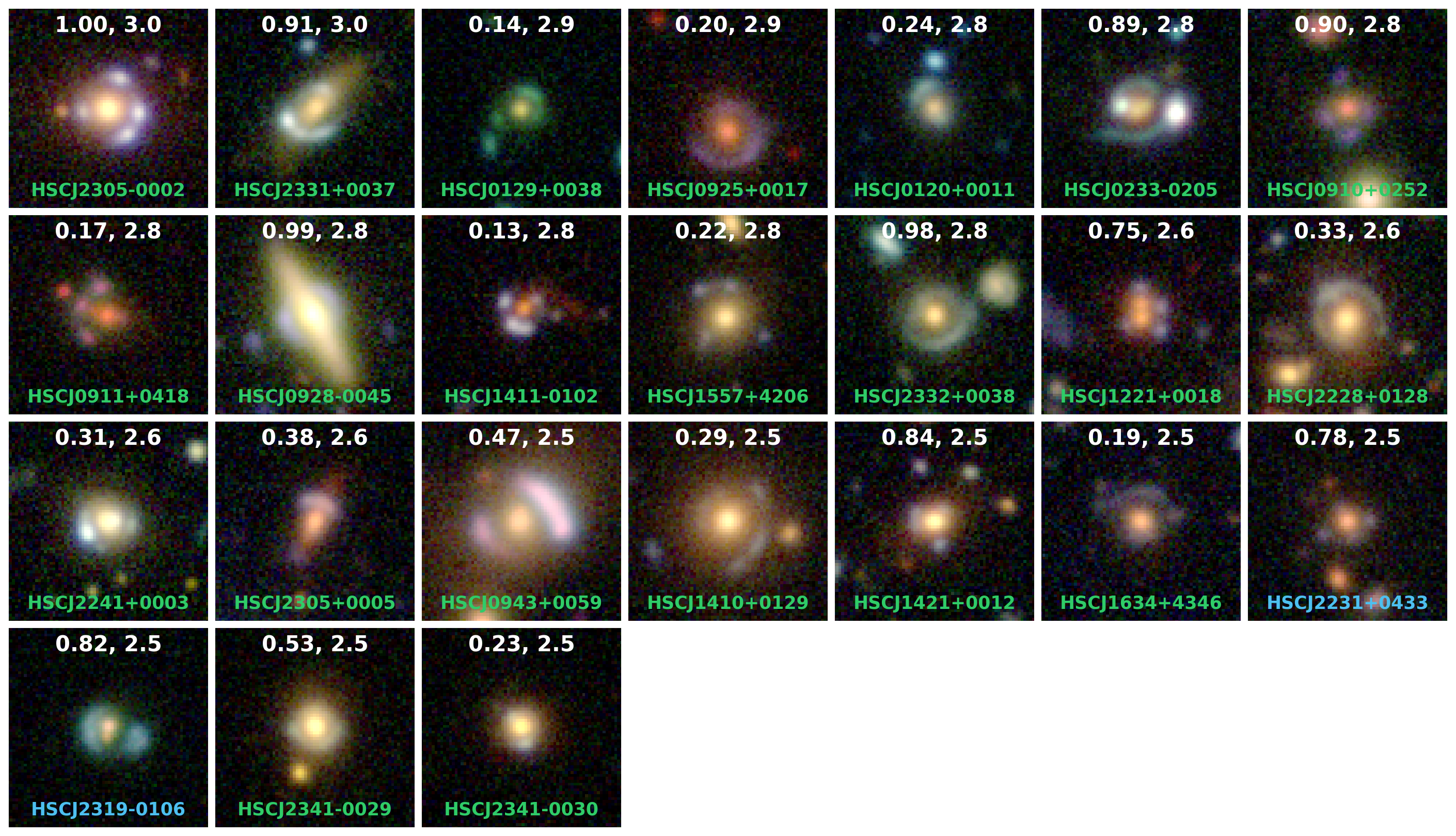}
  \caption{Recentered color-image stamps ($12\arcsec \times 12\arcsec$; north is up and east is left) of identified grade-A lens candidates using HSC $gri$ multi-band imaging data. At the top of each panel, we list the ResNet scores, $p$, and the average grades, $G$, of eight graders, where $\geq 2.5$ corresponds to grade A, from the visual inspection. At the bottom, we list the candidate name, displayed in white for new candidates, light blue if previously detected as grade C lens candidate, and green if previously known grade A or B lens candidate. Given the numerous lens search projects exploiting HSC data, only our grade-B systems (shown in Fig.~\ref{fig:gradeB}) include new identification highlighted additionally with orange boxes. All systems with their coordinates and further details such as the lens environment are listed in Table~\ref{tab:newcand}.}
  \label{fig:gradeA1}
\end{figure*}

Finally, we computed the average grade for all 1 475 systems, where grades from earlier rounds are superseded by newer grades regardless of their value. As in previous lens search projects \citep[e.g.,][]{canameras20, shu22}, we defined grade A lens candidates as systems with final average grade of $G\geq 2.5$ and grade B systems with $2.5 > G \geq 1.5$. In total, we identified finally \newAcand grade-A candidates and \newBcand grade-B candidates. For 75 out of 162 lens candidates, we further correct manually the lens center, as we identified a significant offset here. This is crucial to correctly identify and monitor these lenses in the future. These candidates were selected often through an arc falling into our $i$-Kron range, such that the possible lens galaxy may have a higher $i$-Kron radius. All, if needed recentered, lens candidates are shown in Fig.~\ref{fig:gradeA1} and Fig.~\ref{fig:gradeB}, and listed in Table~\ref{tab:newcand}. The absolute number of identified systems is slightly lower than obtained in previous programs \citep[e.g.,][see also Table~\ref{tab:statisticsoverview} for comparison to \citetalias{canameras21b}]{jacobs19, shu22}. However, they are still comparable with network based searches \citep[e.g.,][]{rojas22, savary22} and depend on the used imaging survey and its footprint size as well. This is most likely due to the sample selection of low $i$-Kron radii systems, containing galaxies with lower mass, consequently less likely to act as strong lenses or creating overly small Einstein radii to be identified with the resolution from HSC (0.168\arcsec/pixel). While this is expected, it is also confirmed by the comparison to \citetalias{canameras21b}, as we can see from Table~\ref{tab:statisticsoverview}.

\subsection{Comparison with known systems}
\label{sec:inspection:known}

From our lens candidates listed in Table~\ref{tab:newcand}, \recovcand candidates are already known based on the current SLED (C. Lemon, private communication, May 2024) and our HOLISMOKES compilation \citep{suyu20}. From these systems, 12 and 60 from our grade A and B sample, respectively, match the grade from the literature. In contrast, 12 and 21 systems from our grade A and B sample, respectively, obtained higher grades than before. On the other hand, 8 published systems with grade A belong to the grade B class according to our grading. We missed a further 62 systems during the visual inspection (i.e., these obtained an average grade $G\leq 1.5$ from us) of the 1475 network candidates, of which 43 have been published with a grade of C (mostly from the SuGOHI sample). Missing grade C candidates is expected as we compile only for grade A and B systems. In total, this shows that a strict grading is difficult as it also depends on the inspected image quality and resolution, filter amount, and possibly detection algorithm (especially for non-DL techniques such as modeling). However, it shows that the expectation on the lensing features are broadly consistent over the years and grading teams.

The relatively high number of known systems is expected since the HSC data set was targeted by multiple lens search projects, complemented by several other lens searches with data in the same footprint. Therefore, to reduce the number of systems during visual inspection, we propose to exclude previously graded systems in the future, regardless of their previous grade, unless the target sample, the data quality (e.g., high-resolution images from Euclid), or the detection algorithm (e.g., including lens deblending or modeling) changed significantly. However, a downside of excluding previously visually classified objects is that we cannot carry out a consistency check with grades from the literature as done above.

\section{Environment analysis}
\label{sec:environment}

Since the surrounding mass distribution of a galaxy-scale lens influences the lensing effect of that system, the characteristics of the environment are crucial and need to be taken into account when building a strong-lens mass model. In addition, galaxy-cluster lenses are modeled completely differently from galaxy-scale lenses and are particularly difficult to identify with autonomous algorithm because of their size, complexity, and peculiarity. However, cluster lenses have several advantages over galaxy-scale lenses, such as significantly longer time delays and higher magnifications, enabling several complementary studies to galaxy-scale systems. For these reasons, beside enlarging the sample to which we apply the network, we conduct an analysis of the lens candidate environment, which we describe in the following. 

Here, we also re-considered our detected lens candidates from \citetalias{canameras21b} and combined them with our new sample. After excluding duplicates, this leads to a total of 546 grade A or grade B lens candidates, based on our visual inspection described in Sect.~\ref{sec:inspection} and \citetalias{canameras21b}.

\subsection{Comparison with the literature}
\label{sec:environment:literature}

As a first step, we cross-matched our network candidates with a large sample of galaxy clusters without known strong-lensing features from the literature. In detail, we first used the catalog from \citet{oguri14}, containing 71 743 clusters identified by the Camira code using data from the SDSS Data Release 8 \citep{york00, aihara11_SDSS8} covering $\sim$11960 deg$^2$ of the sky, including the whole footprint of HSC-SSP DPR2 and thus all our lens candidates. This algorithm is based on stellar population synthesis models to predict colors of red sequence galaxies at a given redshift. The identified clusters cover a redshift range between 0.1 and 0.6 \citep{oguri14}. This code was also directly applied to HSC data, resulting in the 1 921 galaxy clusters presented by \citet{oguri18}. Since the HSC images are significantly deeper than those from SDSS, this catalog extend the redshift range up to $z\sim1$.

We complemented both Camira catalogs with 1 959 galaxy clusters detected by \citet{wen18} using SDSS and Wide-field Infrared Survey Explorer (WISE) data, along with 21 661 galaxy clusters published by \citet{wen21} exploiting HSC-SSP and WISE data. \citet{wen18, wen21} specifically targeted the high redshift range $z\geq 0.7$, complementing ideally identifications by the Camira algorithm. The identifications in \citet{wen21} are based on photometric redshifts obtained from the seven-band photometric data from HSC and SDSS WISE using a nearest-neighbor algorithm. 

This sums up to a total of more than 100 000 galaxy clusters with most systems in the targeted HSC Wide area. Thanks to the two complementary techniques, it also covers a broad redshift range, ensuring a roughly equal distribution over the HSC footprint and reduced selection biases.

The cross-match radius was selected based on the size of known lensing clusters. For instance, \citet{schuldt24} presented 308 securely identified cluster members of the lensing cluster MACS J1149.5+2223, one of the largest sample of cluster members. They are distributed over an area of around 160\arcsec on a side, mostly limited by the area of available high-resolution imaging data. Therefore, to also include systems with significant offset to the BCG, we adopt a radius of 100\arcsec. This size corresponds to $\sim 0.65$ Mpc at the lens cluster redshift of MACS J1149.5+2223 and roughly to the peak of the distribution of the reported R500 radii in \citet{wen21}. The distribution of the reported R500 radii starts at around 0.36 Mpc, which corresponds to 100\arcsec\ at a redshift of 0.25, and the distribution drops significantly until $\sim 0.8$ Mpc, which is equal to 100\arcsec\ at $z=0.9$, supporting our selected radius of 100\arcsec.

In this way, we were able to identify 174 grade A or B lens candidates to be located within 100\arcsec\ from a galaxy cluster centroid listed in the galaxy cluster catalogs. Out of these 174 systems, 63 systems are more than $50\arcsec$ away from the reported cluster center, and only 44 are part of the SuGOHI group- or cluster scale lens sample. We indicate the possible overdense environment from the literature with OD$_\text{lit}$ in Table~\ref{tab:newcand} and Table~\ref{tab:clustercandC21} for the newly identified systems and those from \citetalias{canameras21b}, respectively. These cluster identifications are used in the following as a reference. 

\begin{table*}[t!]
    \caption{High-confidence lens candidates reported by \citetalias{canameras21b} but not listed in Table~\ref{tab:newcand}. An analysis of their environment is now included.}
    \begin{center}
    \begin{tabular}{c|cc|cccc|ccc|ccc|c}
Name   &RA         &Dec       & $p$ & $G$ & $\sigma_\text{G}$ & $z$ & OD$_\text{lit}$ & OD$_\text{vis}$ & OD$_\text{z}$ & $N_\text{max}$ & $z_\text{low}$ & $N_\text{tot}$ & References\\
(1)& (2) & (3) & (4) & (5) & (6) & (7) & (8) & (9) & (10) & (11) & (12) & (13) & (14) \\ \hline \hline \noalign{\smallskip}
HSCJ1004$-$0031 & $151.21577$ & $-0.52915$ & $0.17$ & $3.00$ & $0.00$ & $ 1.05$ & Y & Y & N & 17 & 0.50 & 513 & C21 A23 \\
HSCJ1224$-$0042 & $186.20954$ & $0.704233$ & $0.28$ & $3.00$ & $0.00$ & $ 0.38$ & Y & Y & Y & 23 & 0.50 & 791 &  J20 P19 C21 \\
HSCJ1434$-$0056 & $218.72664$ & $-0.94963$ & $0.94$ & $3.00$ & $0.00$ & $ 0.76$ & Y & Y & Y & 27 & 0.72 & 820 & S18 J20 C21 J23 \\
HSCJ0102$+$0158 & $ 15.65975$ &  $1.98276$ & $0.12$ & $3.00$ & $0.00$ &     & N & N & N & 10 & 0.74 & 259 & C21 A23 \\
HSCJ0238$-$0545 & $ 39.57340$ & $-5.76603$ & $0.72$ & $3.00$ & $0.00$ & $ 1.77$ & N & N & N &  8 & 0.86 & 357 & S18 C21\\
 \vdots & \vdots & \vdots & \vdots & \vdots &  \vdots &  \vdots &  \vdots &  \vdots &  \vdots &  \vdots &  \vdots &  \vdots &  \vdots\\ \hline
    \end{tabular}
    \end{center}
\textbf{Note.} Column description: (1) Source name, (2) right ascension (J2000), (3) declination (J2000), (4) network score, (5) average grade from our visual inspection, (6) dispersion among the eight graders, (7) photometric redshift from our combined catalog based on DEmP \citep{hsieh14}, Mizuki \citep{tanaka18}, and NetZ \citep{schuldt21b}, flag on the overdensity (OD) based on (8) galaxy cluster catalogs \citep{oguri14, oguri18, wen18, wen21}, (9) visual inspection, and (10) photometric redshift, (11) absolute height of photo-$z$ distribution, (12) photo-$z$ lower bound, (13) total number of photo-$z$ in considered area, and (14) references of their previous discovery, apart from \citepalias{canameras21b}, with 
S13 for \citet{sonnenfeld13},
G14 for \citet{gavazzi14},
M16 for \citet{more16b},
D17 for \citet{diehl17},
S18 for \citet{sonnenfeld18a}, 
W18 for \citet{wong18}, 
H19 for \citet{huang19},
P19 for \citet{petrillo19b},
Ch20 for \citet{chan20}, 
Ca20 for \citet{cao20}
L20 for \citet{li20},
J20 for \citet{jaelani20b}, 
S20 for \citet{sonnenfeld20}, 
C21 for \citet{canameras21b},
T21 for \citet{talbot21},
R22 for \citet{rojas22}, 
S22 for \citet{shu22},
A23 for \citet{andika23},
J23 for \citet{jaelani23}, 
and ML for the master lens catalog \url{http://admin.masterlens.org}.
\\
    \label{tab:clustercandC21}
\end{table*}

\subsection{By visual inspection of the environment}
\label{sec:environment:inspection}

\begin{figure*}
\centering
  \includegraphics[trim=0 0 0 0, clip, width=\textwidth]{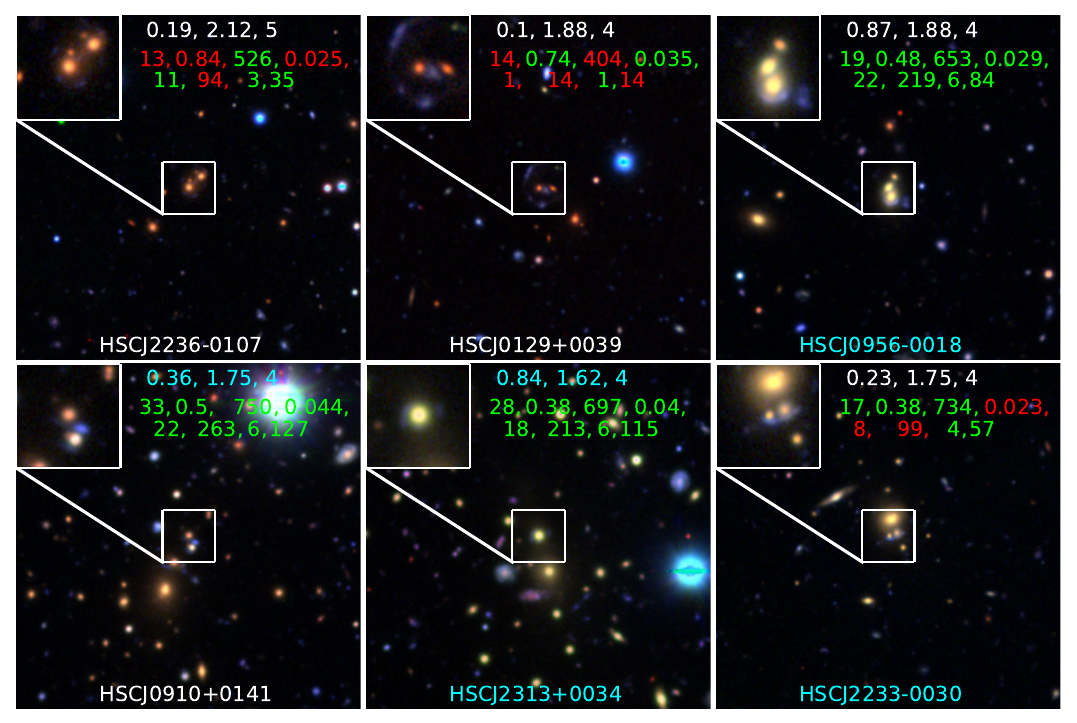}
  \caption{New grade A or grade B lens candidates visually identified to be on group- or cluster-scale, excluding known group- or cluster lenses from the SuGOHI sample. Each panel is $80''\times80''$, and the insert on the top-left shows the $12''\times12''$ cutout analysed by the ResNet. The lens name is given in the bottom of each panel, in cyan if listed in the galaxy cluster catalogs, and otherwise white. Furthermore, the network score, average grade, and number of cluster votes during visual inspection is given in the top (first row), in orange if listed by \citetalias{canameras21b}, in cyan if listed in this work, and white if in both. This is followed by the photo-$z$ characteristics $N_\text{max}, z_\text{low}, N_\text{tot}$, and $N_\text{frac}$ (second row), and $N_\text{peak5}$, $A_\text{peak5}$, $N_\text{peak10}$, and $A_\text{peak10}$ (third row), colored green or red if passing or not passing the final limits (No. 46 in Table~\ref{tab:photozstat}), respectively. We note that lens candidates from \citetalias{canameras21b} were only inspected by a single person for the group and cluster environment classifications and, thus, they have all only one vote. The figure continues in Appendix~\ref{sec:appendixB}. For all shown image stamps, north is up and east is left.}
  \label{fig:clusterexamples}
\end{figure*}

As described in Sect.~\ref{sec:inspection}, we visually inspected all the newly detected network candidates. Here, we consider, besides $10\arcsec \times 10\arcsec$ cutouts that are a typical size used for visual inspection, also stamps of $80\arcsec \times 80\arcsec$ in size, displaying the larger environment (see Fig.~\ref{fig:clusterexamples}). Since all galaxies that belong to a galaxy cluster are located at nearly the same redshift (the so-called cluster redshift) and mostly contain galaxies with similar morphologies, they have similar colors in the images, easing the identification of galaxy clusters based on color images. Consequently, large magnitude catalogs were used in the past to help identify galaxy clusters \citep[e.g.,][]{oguri14}. Instead, we introduced a novel possibility to also indicate a cluster environment to our grading tool, which only slightly increases the required human time for the visual inspection, while offering additional crucial information on the lens candidate. To simplify the identification, we have focused on overdensities in general, which means we also report several group-scale lenses as well as lenses with high number of line-of-sight companions that might not be physically associated with the lens candidate. As a consequence, and analogously to the observed differences in the grades, we also observe some differences in the votes. In the rare case that only one or two graders (out of four or eight, see Sect.~\ref{sec:inspection}) indicated an overdensity, the lens candidate was inspected once more with a focus on the environment for a final decision. Lens candidates with more than two votes were directly indicated as being in a cluster-like environment based on the visual inspection. This is noted as OD$_\text{vis}$ in Table~\ref{tab:newcand} and Table~\ref{tab:clustercandC21}.

Furthermore, one person re-inspected the network candidates reported in \citetalias{canameras21b} regarding the environment, while we adopted the assigned average grades $G$ on the lensing nature from \citetalias{canameras21b} directly. This leads to an identification of additionally 47 (out of 467 grade A or B lens candidates in \citetalias{canameras21b}) to be in a significantly overdense environment, which are listed in Table~\ref{tab:clustercandC21}. By combining both samples and removing duplicates, among the new candidates from Sect.~\ref{sec:inspection} and those from \citetalias{canameras21b} that we visually identified, out of 546 grade A or B lens candidates, 84 are found to be in an overdense environment. Interestingly, from these 84 lens candidates, only 31 were reported in the SuGOHI group- and cluster-scale sample. This demonstrates the necessity of further analysis of the lens environment to enlarge the sample of group- and cluster-scale lenses. In addition, only 54 of the 84 identified systems were reported in the considered galaxy cluster catalogs, and only 27 in both (the SuGOHI and the galaxy cluster catalogs). This highlights that several lens candidates recently identified with deep learning classifiers were lacking in terms of information on their environment. 

\subsection{By photometric redshifts}
\label{sec:environment:phototz}

As a further characterization of the environment from our new lens candidates, as well as the lens candidates from \citetalias{canameras21b}, we obtained for each candidate the photometric redshift (hereafter photo-$z$) distribution within a given area, as detailed below. This analysis goes therefore beyond the analysis of simple magnitude catalogs and will be broadly applicable to upcoming wide field surveys such as Euclid and LSST, with large and accurate photo-$z$ catalogs \citep[e.g.,][]{schmidt20, euclid_X, euclid_XXXI}. We elaborate possible criteria such as the height of the distribution peak and the sum of all objects with available photo-$z$ value in the considered field to identify additional lenses in overdensities. In addition, the peak of the photo-$z$ distribution indicates the cluster redshift and helps to determine whether the lens galaxy belongs to the cluster.

For this analysis, we use photometric redshift catalogs from three different and complementary codes that were broadly applied to objects from the HSC wide area. In detail, these codes are Mizuki \citep{tanaka18}, a template fitting code with Bayesian priors, DEmP \citep{hsieh14}, a hybrid machine-learning code based on polynomial fitting, and NetZ \citep{schuldt21b}, a CNN-based code that infer the photo-$z$ values directly from the HSC $grizy$-image stamps. All three catalogs obtained a very good performance against spectroscopic redshifts, and contain, after excluding duplicates defined by a distance of less than two pixels ($ \leq 0.336\arcsec$), each on the order of several ten million redshifts. In case of duplicates, we took the simple average of all (typically two) photo-$z$ values as well as the coordinates, and considered them as a single object. This ensures that we do not artificially overestimate the density of objects in a given area and expect to reduce the effect of catastrophic outliers. The relatively low threshold of only two pixels is chosen to not introduce wrong photo-$z$ values through averaging. We further follow \citet{schuldt21b}, and limit the catalogs to a photo-$z$ range $0<z<5$, given that the trustworthiness of photo-$z$ values significantly decreases with increasing redshift. 
By combining the three catalogs, we ultimately obtained a photo-$z$ catalog containing more than 115 million redshifts in the targeted footprint area.

Based on the size of known lensing clusters (compare with Sect.~\ref{sec:environment:literature}), we considered all objects within a $200\arcsec$ $\times$ $200\arcsec$ field, centered at the lens candidate in the analysis. Since the cluster member galaxies are typically within a redshift bin of around 0.03 to 0.06 \citep[see e.g.,][]{bergamini21, bergamini23a, acebron22a, acebron22b, schuldt24}, we created a photo-$z$ histogram for all lens candidates with a redshift bin width of $0.02$. However, we note that the range of the cluster members in our distributions is broadened given the photo-$z$ uncertainties. We show the histograms of two lens candidates, one visually identified to be in an overdense environment and one not, in Fig.~\ref{fig:hist_examples} as example.


For the selection of overdensities, we first exploited the extracted photo-$z$ distributions directly and defined the following eight quantities that we used for the evaluation process:

\begin{enumerate}
    \item[$\bullet$] $N_\text{max}$: Since a higher peak of the photo-$z$ distribution indicates a higher concentration of galaxies at a similar redshift, we introduce the absolute height of the photo-$z$ distribution peak $N_\text{max}$ as a criterion.
    \item[$\bullet$] $N_\text{tot}$: The second criterion is the total amount of objects in the adopted field, indicating the density of the field in general.
    \item[$\bullet$] $N_\text{frac}$: The third criterion is the ratio $N_\text{max}/N_\text{tot}$, indicating the concentration of systems at the given redshift bin compared to the whole distribution.
    \item[$\bullet$] $N_\text{peak5}$: Since, as noted above, we would expect a galaxy cluster to cover multiple neighboring redshift bins, we further introduce a criterion that gives the number of bins next to $N_\text{max}$ exceeding five counts.
    \item[$\bullet$] $N_\text{peak10}$: Same criterion as $N_\text{peak5}$ but for 10 neighboring bins instead of five.
    \item[$\bullet$] $A_\text{peak5}$: This criterion gives the sum of objects within $N_\text{peak5}$ bins, since a higher number of systems within the bounds above five indicate a significant overdensity.
    \item[$\bullet$] $A_\text{peak10}$: This criterion gives, in analogy to $A_\text{peak5}$, the sum of objects within $N_\text{peak10}$ bins.
    \item[$\bullet$] $z_\text{low}$: The lower redshift bound value of $N_\text{max}$.
\end{enumerate}

The first seven criteria rely on the fact that the different and complementary photometric redshift algorithms were applied broadly to the same footprint of the network candidates. These criteria are highlighted in the photo-$z$ histograms shown in Fig.~\ref{fig:hist_examples}. The criterion $z_\text{low}$ was introduced since, based on the SuGOHI sample, galaxy clusters at very low redshift ($z \lesssim 0.1$) or very high redshift ($z \gtrsim 1$) are very unlikely to create strong-lensing effects detectable in the HSC images that make up our target sample. In the case where two redshift bins have the exact same number of systems, $N_\text{max}$, we used the lower redshift bin value as $z_\text{low}$ since higher photometric redshifts are normally less accurate; consequently, a peak at higher redshift might be just due to photo-$z$ outliers rather than an actual galaxy cluster. We note that this choice does not affect our first three criteria at all and that a combination of these criteria is crucial, as detailed below.

Since we are only interested in the characterization of strong-lensing systems, and because galaxy-scale lenses are in general located in fields with higher density \citep[e.g.,][]{wells24}, we limited the analysis to our 546 visually identified lens candidates. Nonetheless, we ran the photo-$z$ analysis on the whole set of network candidates with $p \geq 0.1$, as well as 10,000 random positions in the HSC footprint, for comparison and consistency checks. Based on the considered galaxy cluster catalogs and the visual inspection, this sample of 546 lens candidates contains, 174 and 84 systems, respectively, in a significantly overdense environment. 

\begin{figure*}
\centering
  \includegraphics[trim=0 0 0 0, clip, width=\textwidth]{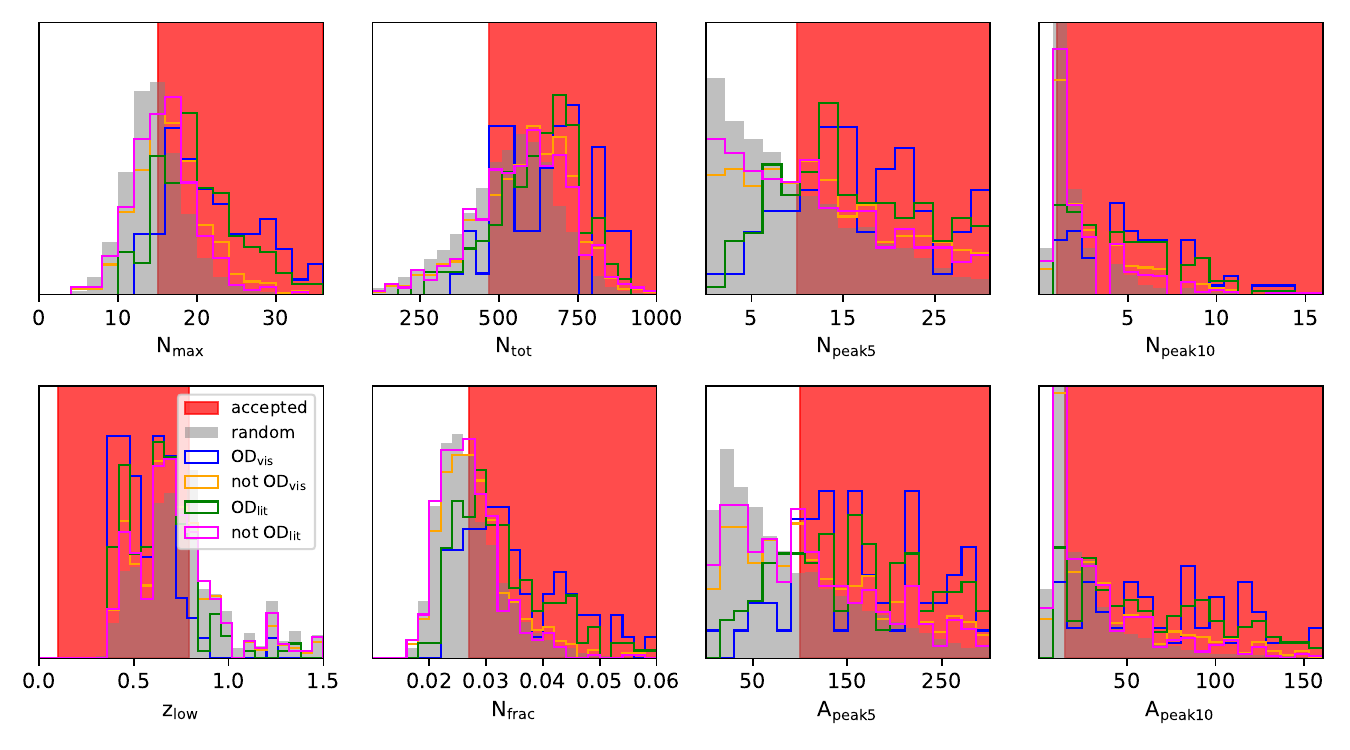}
  \caption{Normalized histograms of our eight introduced selection criteria. We distinguish between lens candidates visually identified to be in an overdensity (blue) or through the galaxy cluster catalogs (green), compared to those not in an overdensity (orange and magenta, respectively). We further show, for comparison, the distribution from 10,000 random positions in gray. Following the F1 criterion, we highlight the parameter range indicating overdensities in red (shaded regions for criteria No.~46 in Table~\ref{tab:photozstat}), which demonstrates that a combination of these different criteria is crucial to gain good performance (see also No. 50 to 54 in Table~\ref{tab:photozstat}).}
  \label{fig:photozcomparison_orighist}
\end{figure*}

We show in Fig.~\ref{fig:photozcomparison_orighist} the distribution obtained for our eight introduced selection criteria. This highlights the difference between systems in overdensities (blue and green) compared to those in the field (orange and magenta), and consequently the effectiveness of our selection criteria. For comparison, we show in gray the inferred values from 10,000 random positions in the HSC footprint. This confirms also findings from \citet{wells24} that lenses are in general in overdense environments.

We then set different thresholds for these eight criteria and tested them against the visually identified lens candidates. While it is possible that we missed some clusters during our visual inspection, and we acknowledge the considered galaxy cluster catalogs may be incomplete, we treated these samples as the ground truth for inferring the best thresholds for our eight introduced criteria. This allowed us to define the true-positive (TP) and false-positive (FP) rates, which we listed for a representative selection of different limits in Table~\ref{tab:photozstat}. For comparison, we included, as No. 0, the full sample sizes of lens candidates (indicated by selection cuts that all candidates passes the criteria by definition).

Finally, we used the F1 criterion defined as
\be
\text{F1} = \frac{2 \times \text{precision} \times \text{recall}}{\text{precision} + \text{recall}}~,
\ee
with
\be
\text{precison} = \frac{\text{TP}}{\text{TP+FP}}
\ee
and
\be
\text{recall} = \text{TPR} = \frac{\text{TP}}{\text{P}} ~,
\ee
where P denotes the positive sample size (lens candidates in overdensity) to identify the best selection cuts. We list in Table~\ref{tab:photozstat} the F1 values using the performance on the visually classification, denoted as F1$_\text{vis}$, the cluster catalogs, denoted as F1$_\text{lit}$, and their combination, denoted as F1$_\text{tot}$.

Following the F1$_\text{tot}$ criterion, No. 46 shows the best performance. It also has the highest score for F1$_\text{vis}$, while slightly outperformed according to F1$_\text{lit}$. These selection cuts are indicated in Fig.~\ref{fig:photozcomparison_orighist} in red. 

We test the individual selection cuts from No. 46 to see their selection effect (No. 50 to 54 in Table~\ref{tab:photozstat}), while keeping the cut of $z_\text{low} \geq 0.1$ as this criterion rejects photo-$z$ outliers rather than non-overdensities and is thus applied for all tested combinations. This clearly highlights the importance of the combination of multiple selection cuts to obtain a good performance.  

Since the distribution of objects in a random field is not flat with respect to redshift, we also test our selection criteria using subtracted photo-$z$ histograms. This means we created the photo-$z$ histograms for 10 000 random positions in the HSC footprint using our compiled photo-$z$ catalog, and computed the median and the standard deviation per redshift bin. To obtain better statistics at these random positions, we went on to consider a bin width of 0.06. We then subtracted the median histogram from that of all our lens candidates and applied our criteria to the remaining distribution. As an alternative, we also tested normalized histograms, defined as the subtracted histograms divided by the standard deviation obtained from the 10 000 histograms at random positions. We show the obtained distribution of our eight selection criteria in Fig.~\ref{fig:photozcomparison_subtrhist} and Fig.~\ref{fig:photozcomparison_normhist}, respectively. A compilation of possible cuts using the subtracted histograms are listed in Table~\ref{tab:photozstat_subtr}. According to the F1 criterion, the selection on the original histograms shows a better performance. This might be because of the relatively large considered area of $200\arcsec \times 200\arcsec$, centered at the lens candidate, consequently including multiple foreground or background galaxies. Thus, although we subtracted the median of 10 000 random positions, compact galaxy groups do not show a prominent excess in the photo-$z$ distribution of their surrounding objects.

\subsection{Final cluster selection}
\label{sec:environment:final}

Based on our comparison of different cuts on our eight introduced criteria, and taking into account the completeness and purity of the selection through the F1 criterion, we favor No.~46 in Table~\ref{tab:photozstat}. This gives a completeness rate on the visually identified systems of 58/84 $\sim$ 70\%, while 87/174 =50\% are listed in the considered galaxy cluster catalogs. We recall that we flagged all overdensities during our visual inspection, including possible group-scale lenses and possible overdense fields not necessarily with most systems at a similar redshift. On the other hand, since the primary focus was the lens grading, the sample cannot be considered as complete. This is already evident from the considered catalogs, where we found 174 lens candidates less than 100\arcsec\ away from a known galaxy cluster, although it is expected that the most distant matches might be missed as we inspected only $80\arcsec \times 80\arcsec$ cutouts centered at the lens candidate. Therefore, it is understandable that we identified some candidates with our photo-$z$ selection that had been missed either during the visual inspection or by the galaxy cluster catalogs, which are also incomplete. In contrast, the photo-$z$ selection misses some candidates given their distance to the galaxy cluster or due to missing photo-$z$ measurements around the lens candidate. For instance, this is the case for HSC J2329-0120, a grade-B lens candidate visually identified as cluster lens and listed in the galaxy cluster catalog from \citet{wen21}, as well as the SuGOHI group- and cluster-scale lens sample. While the image stamps analyzed by the network are perfectly fine in all three bands, the $g$ and $r$ bands show $\sim 14\arcsec $ to the north a $~19\arcsec$ broad stripe without observations. Consequently, in that area, no photometric redshifts are available, resulting in a lower number of galaxies in the field. Thus, it did not pass with a value of 395, our restriction on $N_\text{tot}$, while fulfilling the other criteria. These aspects help explain the missing lens candidates identified as lens clusters with other techniques. On the other hand, with selection No. 46, we found less than 15\% contamination, namely lens candidates not identified as galaxy cluster that have nonetheless passed our photo-$z$ criteria.

In total, we have 58 visually identified lens clusters that pass our criteria on the photo-$z$ limits, which include 22 known systems from the SuGOHI group- and cluster-lens catalogs. From these 58 lens cluster candidates, only 15 are not listed in the galaxy cluster catalogs, highlighting the efficiency of the visual inspection in combination with our photo-$z$ criteria. This implies that we identified 43 lens candidates that are (1) listed in the galaxy cluster catalogs, (2) visually identified as clusters, and (3) passing our photo-$z$ criteria, while not being reported in the SuGOHI group- and cluster-scale sample. While these are the most secure newly identified cluster-lens candidates, we considered all lens candidates that pass one of these three criteria as a lens candidate in a significantly overdense environment, resulting in a total of 231/546 systems. For comparison, 136 of 546 lens candidates and 879 random fields from our 10 000 comparison sample, respectively, do indeed pass our photo-$z$ criteria. This shows once more that lenses tend to be in an overdense environment compared to random galaxies, in agreement with the findings of e.g., \citet{wells24}. 

\section{Mass models and their statistical analyses}
\label{sec:massmodel}

As a further step of the lens candidate analysis, we exploited the residual neural network presented by \citet{schuldt23a} to model the mass distribution of all 546 grade A and B lens candidates, making this the largest sample that has been modeled in an uniform way and the first such statistical analysis overall. However, we note that this is only a first step towards a statistical analysis of a whole lens population and a detailed analysis is relegated to a further work.

\begin{figure*}
\centering
  \includegraphics[trim=120 20 100 50, clip, width=\textwidth]{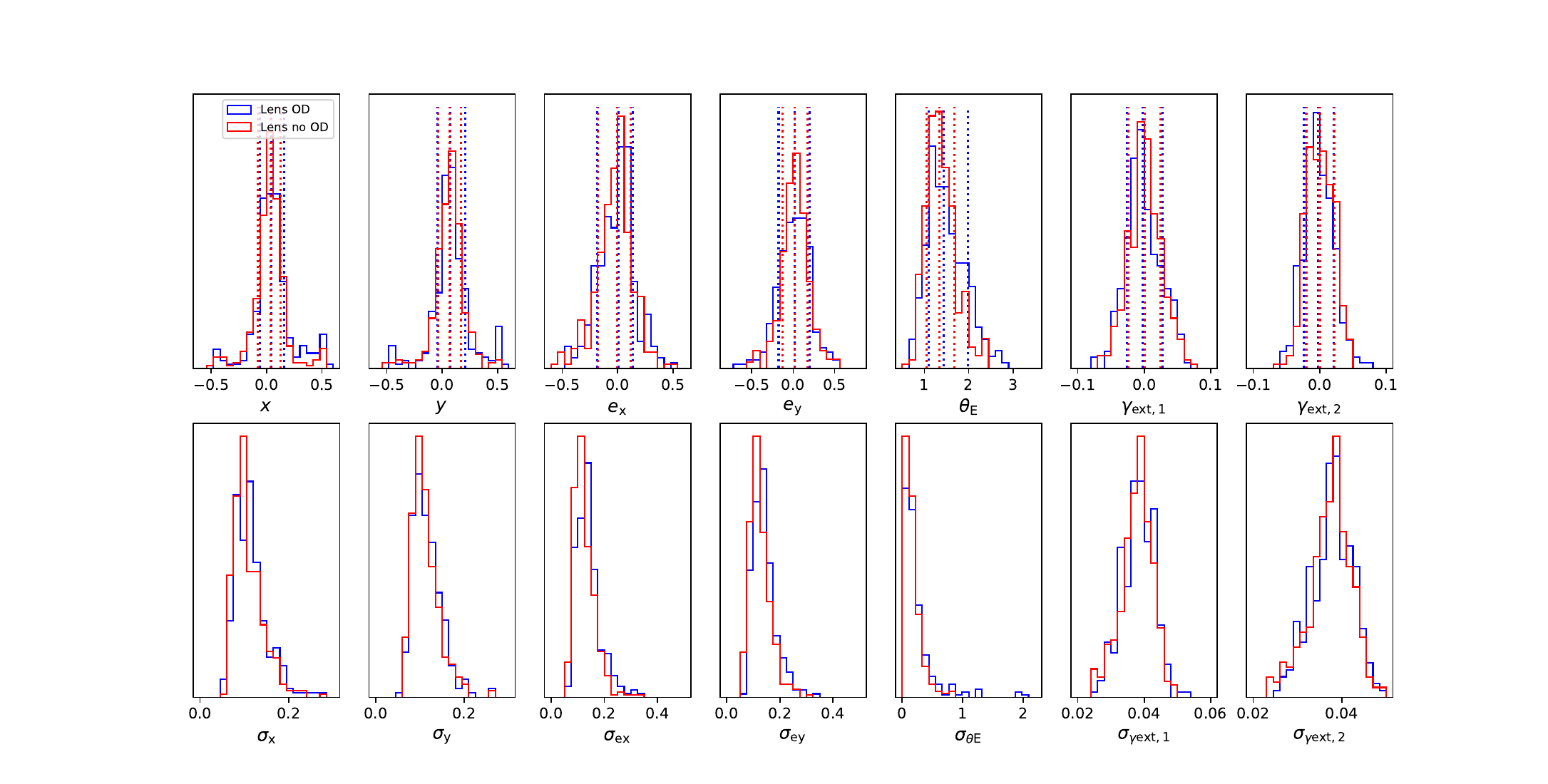}
  \caption{Normalized histograms of the mass model parameter values (top row) with uncertainties (bottom row) predicted by the neural network of \citet{schuldt23a}. From left to right: Predict values for the SIE profile (lens center, $x$ and $y$, complex ellipticity, $e_\text{x}$ and $e_\text{y}$, and Einstein radius, $\theta_\text{E}$) and the external shear ($\gamma_\text{ext,1}$ and $\gamma_\text{ext,2}$). We applied the network to all grade A and B lens candidates identified in \citetalias{canameras21b} and this work, and distinguish between those flagged by any overdensity criteria (blue) and those not (red). The median values and $\pm 1$ sigma ranges are indicated by dotted vertical lines for each parameter (see also Table~\ref{tab:modelpar}).}
  \label{fig:modelparametercomparison}
\end{figure*}

The network has been trained on highly realistic HSC-based mock lens images generated with the same code as the mocks used in \citetalias{canameras21b} to train the lens finding network that we exploited in this work (see Sect.~\ref{sec:methodology}). It predicts the mass model parameter values of a SIE mass profile with external shear and was tested on 31 confirmed HSC lenses, resulting in a good recovery for most of the mass model parameters, except the external shear \citep{schuldt23b}. The distribution of the estimated parameter values of our 546 lens systems is shown in the top row of Fig.~\ref{fig:modelparametercomparison}, while the predicted uncertainties at the bottom. We give in Table~\ref{tab:modelpar} the median with $1\sigma$ values for the seven predicted parameters, which are also marked by vertical lines in Fig.~\ref{fig:modelparametercomparison}. Here, we distinguish among the lens candidates flagged by any of our three criteria to identify an overdensity (literature, visual inspection, or photometric redshift) and those that are not flagged. Therefore, the overdensity sample includes also some group-scale lenses and some FPs, namely, galaxy-scale lenses (see also Table~\ref{tab:photozstat}), biasing the statistics towards the system of galaxy-scale lenses in the field. We further note that the modeling residual neural network was only trained for galaxy-scale systems and, thus, it is forced to predict an Einstein radius value between 0.5 and 5\arcsec, for instance.

{\renewcommand{\arraystretch}{1.2}
\begin{table}[ht!]
    \caption{Median with their $1\sigma$ (86\% percentile) values of the seven parameters predicted by the network of \citet{schuldt23a}.\label{tab:modelpar}}
    \begin{center}
    \begin{tabular}{c|cc}
   Parameter & with OD & without OD \\ \hline \hline \noalign{\smallskip}
$x$ & $ 0.05 ^{+0.12}_{- 0.11} $ & $ 0.04 ^{+0.10}_{-0.12} $ \\
$y$ & $ 0.07 ^{+0.14}_{- 0.12} $ & $ 0.07 ^{+0.11}_{-0.11} $ \\
$e_\text{x}$ & $ 0.01 ^{+0.13 }_{- 0.19 } $ & $-0.01 ^{+0.13 }_{- 0.17 } $ \\
$e_\text{y}$ & $ 0.02 ^{+0.18 }_{- 0.20 } $ & $ 0.02 ^{+0.16 }_{- 0.15 } $ \\
$\theta_\text{E}$ & $ 1.43 ^{+0.55 }_{- 0.34 } $ & $ 1.33 ^{+0.35 }_{- 0.28 } $ \\
$\gamma_\text{ext,1}$ & $-0.00 ^{+0.03 }_{- 0.03 } $ & $ 0.00 ^{+0.03 }_{- 0.03 } $ \\
$\gamma_\text{ext,2}$ & $-0.00 ^{+0.03 }_{- 0.03 } $ & $ 0.00 ^{+0.03 }_{- 0.02 } $\\
    \end{tabular}
    \end{center}
    \textbf{Notes.} We separated the systems into samples flagged by any of our three overdensity (OD) criteria or those that did not get flagged at all. The corresponding histograms are shown in Fig.~\ref{fig:modelparametercomparison}. 
\end{table}
}

As expected, the predicted Einstein radii in an overdensity tend towards higher values (see particularly the $+1\sigma$ bound), while being still compatible with galaxy-scale systems. We did not identify a significant difference for the other parameters, nor the predicted uncertainties. While we would expect a tendency towards higher external shear for systems in an overdensity, we argue that the consistency is due to the difficulties with this network to predict the external shear. This is a known issue \citep[see][]{schuldt23a, schuldt23b} and likely a result of the image resolution and seeing in HSC images observed from the ground in contrast to high-resolution observations from space (e.g., with \HST, JWST, or Euclid).

While we compared above systems flagged by any of our overdensity criteria to systems that did not get flagged, we also performed the comparison with systems that have been identified by all three of our criteria. This results in a very clean sample with only lens candidates in clear overdensities. In this case, we obtained an Einstein radius of $ 1.60 ^{+0.56}_{-0.50} $, which is significantly higher than that for the remaining systems and also notably higher than those reported in Table~\ref{tab:modelpar}. This is in agreement with the expectation that the cluster environment does help to increase the size of the Einstein radius. All the other parameters are still consistent. However, we note that this value is based on a relatively small sample with only 43 systems.

\section{Conclusion and discussion}
\label{sec:conclusion}

We extended our systematic search of galaxy-scale lenses by doubling our considered candidates to $\sim $ 135 million objects observed in the HSC Wide survey. We exploited the well tested residual neural network presented by \citetalias{canameras21b} and obtained 11 816 network candidates from the new sample. To identify the most probable lens candidates and reject false positives, we carry out a multi-stage visual inspection with eight individual graders. This ensures stable average grades while limiting the workload though pre-selection with fewer graders. We discuss the procedure in detail, and elaborate recommendations for further visual inspections particularly on large samples such as those expected from the Euclid and \textit{Rubin} LSST surveys. The proposed strategy is as follows:
\begin{itemize}
    \item Calibration round: every grader inspects the same small sample, which is then discussed to agree on the grading criteria. This is particularly crucial for new data sets or grading teams.
    \item Given the high number of lens search projects carried out, we propose excluding all previously classified network candidates (i.e., candidates classified as lenses and non-lenses) to lower the number of systems that need inspection, unless the detection algorithm (e.g., modeling rather than CNNs) or data set (e.g., high resolution images from Euclid) has changed notably.
    \item To reject a significant fraction of false positives, we propose a first inspection by a single person. The goal of this binary classification is to reject only the most obvious interlopers to lower the amount of systems that need to be inspected in the following steps, while keeping any system that require a closer look or system that may gain from a second opinion. In our case, this removed $\sim90\%$ of the network candidates, mostly images with, for instance, artefacts, or bright saturated stars that were not included in the training data.
    \item A group of around four individuals inspect the remaining network candidates, providing grades between 0 and 3, and any additional classification flag included (e.g., indicating an offset of the lens, a cluster environment, a possible lensed quasar). 
    \item To further increase the grades per interesting object, but lowering the amount of human inspection, we propose to only collect additional grades for objects with average grade above 1, to finally average over at least seven grades per relevant object. While the threshold for interesting objects is $G\geq1.5$, we propose here a slightly lower threshold to allow for potential upgrading of candidates after acquiring more grades and to not exclude possible outliers in the grades.
    \item Finally, a re-classification of systems with high-dispersion among the provided grades.
\end{itemize}

Following a similar approach, which is described in Sect.~\ref{sec:inspection}, we present \newAcand grade A lens candidates (average grade $G\geq 2.5$) and \newBcand grade B lens candidates ($2.5>G\geq1.5$) in Table~\ref{tab:newcand}, containing \recovcand recovered lens candidates that had already been identified by complementary algorithms.

While there have been a variety of lens search projects carried out, they are nearly always focusing on static, not time-varying, galaxy-scale lenses. Given the impact of the lens environment in their analysis and the advantages of cluster-scale lenses for various studies as well as the possibility of exploiting them to study galaxy cluster properties, we present a detailed analysis of their environment. Here, we consider also our previously identified network candidates from \citetalias{canameras21b}, enlarging the sample to 546 grade A or B lens candidates. For this, we visually inspected additionally larger image stamps as shown in Fig.~\ref{fig:clusterexamples} and classified their environment. 

We further compiled and exploited a photo-$z$ catalog obtained from three complementary techniques broadly applied to the HSC Wide survey area, ultimately providing more than 115 million photo-$z$ values. The presented criteria enable a selection of cluster-lens candidates tested against our visual lens candidate identification and known galaxy clusters from the literature. 

After their identification and environment analysis, we applied the neural network from \citet{schuldt23a} to all 546 analyzed grade A and B lens candidates, making it the largest sample that has been uniformly modeled so far. We discussed the parameter value and corresponding uncertainty distributions, while considering galaxy-scale systems in an overdensity separately from those in the field. We found a tendency towards larger Einstein radii for systems in an overdensity, while other parameters remain the same for the sample.

The proposed environment analysis and statistical modeling analysis pave the way to efficiently exploit larger samples of lens candidates in the upcoming era of wide-field imaging surveys. From this, we expect around 100~000 lenses, and which will bolster large and accurate photo-$z$ catalogs, as well as offer a novel technique to identify new lensing clusters.

\section*{Data availability}
Tables \ref{tab:newcand} and \ref{tab:clustercandC21} are available in electronic form at the CDS via anonymous ftp to \url{cdsarc.u-strasbg.fr} (130.79.128.5) or via \url{http://cdsweb.u-strasbg.fr/cgi-bin/qcat?J/A+A/}.

\begin{acknowledgements}
We thank C. Lemon for sharing the SLED compilation of known lens candidates and the anonymous referee for the helpful comments. SS has received funding from the European Union’s Horizon 2022 research and innovation programme under the Marie Skłodowska-Curie grant agreement No 101105167 — FASTIDIoUS.
We thank the Max Planck Society for support through the Max Planck Fellowship of SHS. This project has received funding from the European Research Council (ERC) under the European Union’s Horizon 2020 research and innovation programme (LENSNOVA: grant agreement No 771776). This research is supported in part by the Excellence Cluster ORIGINS which is funded by the Deutsche Forschungsgemeinschaft (DFG, German Research Foundation) under Germany's Excellence Strategy -- EXC-2094 -- 390783311. 
SB acknowledges the funding provided by the Alexander von Humboldt Foundation.
CG acknowledges financial support through grants PRIN-MIUR 2017WSCC32 and 2020SKSTHZ.
This paper is based on data collected at the Subaru Telescope
and retrieved from the HSC data archive system, which is operated by Subaru Telescope and Astronomy Data Center at National Astronomical Observatory of Japan. The Hyper Suprime-Cam (HSC) collaboration includes the astronomical communities of Japan and Taiwan, and Princeton University. The HSC instrumentation and software were developed by the National Astronomical Observatory of Japan (NAOJ), the Kavli Institute for the Physics and Mathematics of the Universe (Kavli IPMU), the University of Tokyo, the High Energy Accelerator Research Organization (KEK), the Academia Sinica Institute for Astronomy and Astrophysics in Taiwan (ASIAA), and Princeton University. Funding was contributed by the FIRST program from Japanese Cabinet Office. the Ministry of Education. Culture, Sports, Science and Technology (MEXT), the Japan Society for the Promotion of Science (JSPS), Japan Science and Technology Agency (JST), the Toray Science Foundation, NAOJ, Kavli IPMU, KEK, ASIAA, and Princeton University.
This work uses the following software packages:
\href{https://github.com/astropy/astropy}{\texttt{Astropy}}
\citep{astropy1, astropy2},
\href{https://github.com/matplotlib/matplotlib}{\texttt{matplotlib}}
\citep{matplotlib},
\href{https://github.com/numpy/numpy}{\texttt{NumPy}}
\citep{numpy1, numpy2},
\href{https://www.python.org/}{\texttt{Python}}
\citep{python},
\href{https://github.com/scipy/scipy}{\texttt{Scipy}}
\citep{scipy}.
\href{https://www.star.bris.ac.uk/~mbt/topcat/}{\texttt{TOPCAT}}
\citep{topcat}
\end{acknowledgements}

\bibliographystyle{aa}
\bibliography{main}

\appendix

\section{Overview of grade-B lens candidates}
\label{sec:appendixA}

\begin{figure*}[t!]
  \includegraphics[trim=0 0 0 0, clip, width=0.95\textwidth]{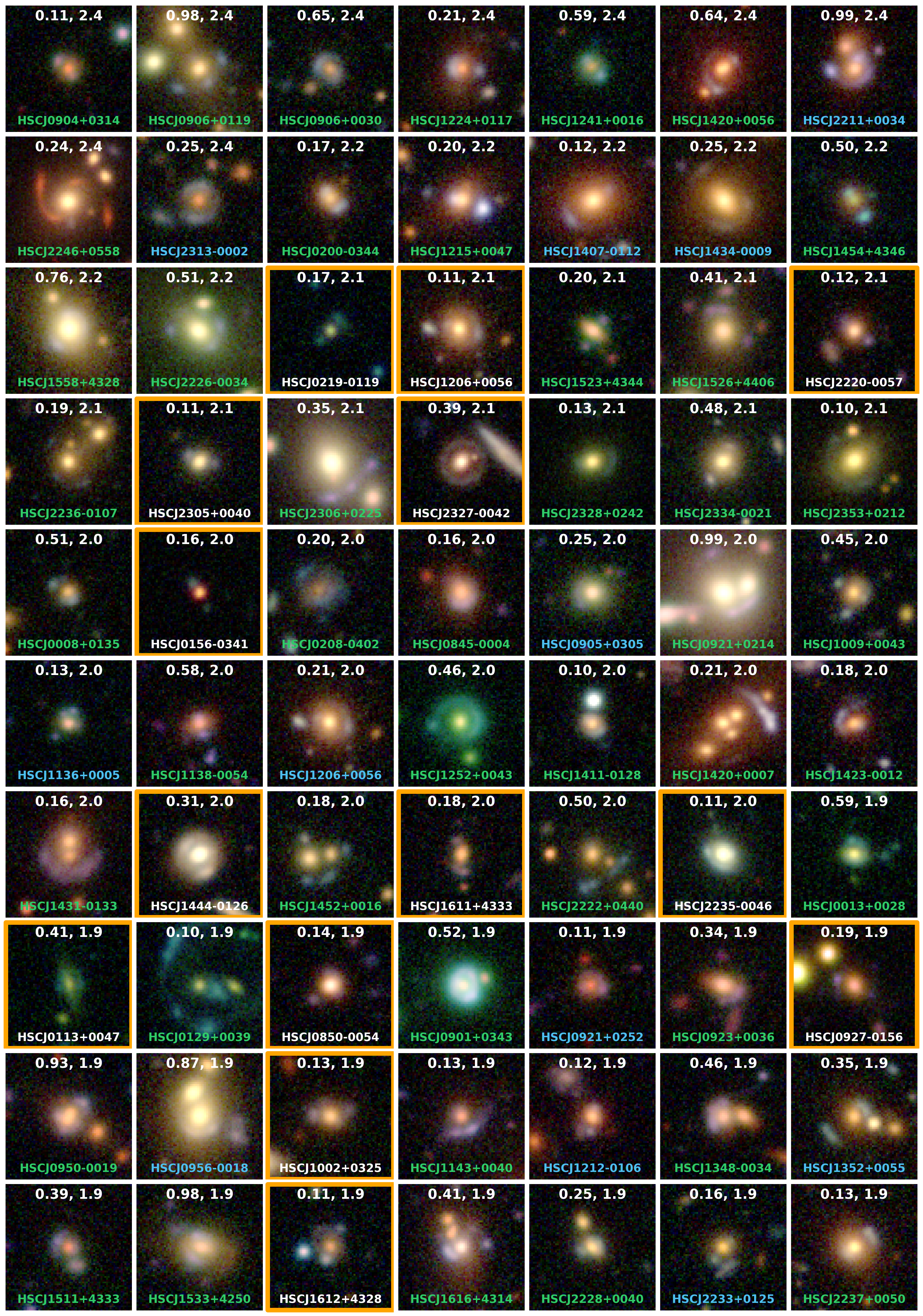}
  \caption{Recentered color-image stamps of identified grade-B lens candidates. Same format as Fig.\ref{fig:gradeA1}.}
  \label{fig:gradeB}
\end{figure*}

In this appendix, we show in Fig.~\ref{fig:gradeB} recentered color-image stamps of our grade B candidates, in analogy to Fig.~\ref{fig:gradeA1}. The image stamps are 12\arcsec\ on a side and we display the corresponding HSC name in the bottom of each panel, appearing white if discovered for the first time, and light blue if the candidate was previously classified as a grade C candidate. Names in green indicate re-discoveries with comparable grade. In the top we provide the network score $p$ and the average grade $G$. Fully new discoveries are further highlighted with orange frames.

\begin{figure*}[t!]
  \includegraphics[trim=0 0 0 0, clip, width=0.95\textwidth]{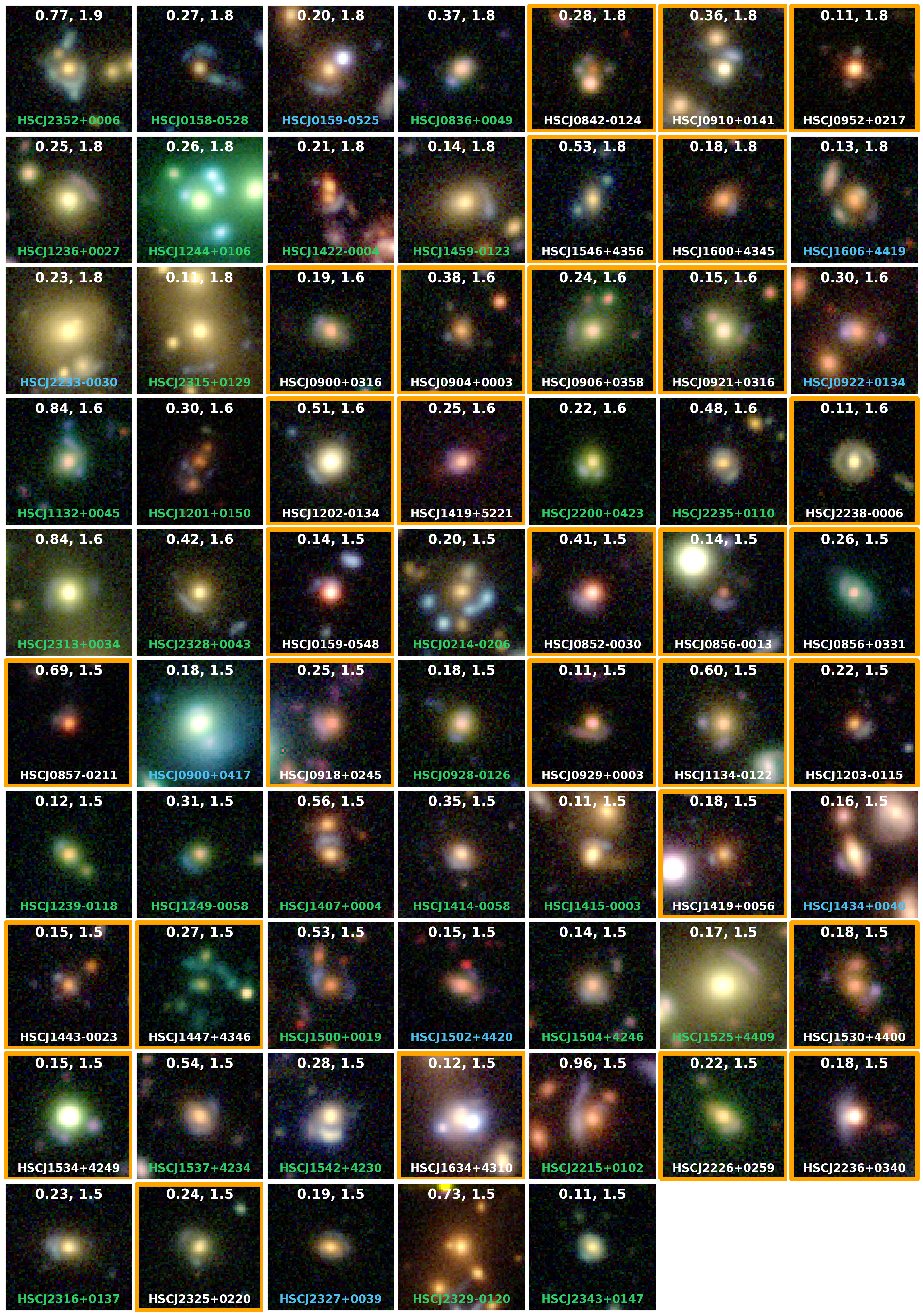}
  \caption*{Fig.~\ref{fig:gradeB} continued: Recentered color-image stamps of identified grade-B lens candidates.}
\end{figure*}

\FloatBarrier
\section{Details of lens candidates in overdensities}
\label{sec:appendixB}

\begin{table*}[p]
    \caption{Comparison of true-positive (TP) and false-positive (FP) rates for different selection criteria using the photometric redshift distributions around the 546 grade A or B lens candidates.}
    \begin{center}
    \begin{tabular}{c|ccccccccc|cc|cc|ccc}
   No. &\multicolumn{9}{c|}{Criteria} & \multicolumn{2}{c|}{visual insp.} & \multicolumn{2}{c|}{Literature} & \multicolumn{3}{c}{F1 statistics}\\ \hline \noalign{\smallskip}
   & $N_\text{max}$ & $z_\text{low}$ & $z_\text{low}$& $N_\text{tot}$ & $N_\text{frac}$ & $N_\text{peak5}$ & $A_\text{peak5}$ & $N_\text{peak10}$ & $A_\text{peak10}$ & TP & FP & TP & FP & F1$_\text{vis}$ & F1$_\text{lit}$ & F1$_\text{tot}$ \\ 
\hline \hline \noalign{\smallskip}
   0 & $\geq  0$ & $\geq 0$ & $\leq \inf $ & $\geq 0$ & $\geq 0$ & $\geq  0$ & $\geq  0$ & $\geq  0$ & $\geq  0$& 84 & 462 & 174 & 372 & 0.210 & 0.483 & 0.358 \\ \hline \hline
   
   1 & $\geq 15$ & $\geq 0.1$ & $\leq 1.0$ & $\geq 400$ & $\geq 0.025$ & $\geq  0$ & $\geq  0$ & $\geq  0$ & $\geq  0$ & 68 & 188 & 119 &137 & 0.425 & 0.553 & 0.499  \\
   2 & $\geq 15$ & $\geq 0.1$ & $\leq 1.0$ & $\geq 400$ & $\geq 0.026$ & $\geq  0$ & $\geq  0$ & $\geq  0$ & $\geq  0$ & 67 & 167 & 115 &119 & 0.450 & 0.564 & 0.516 \\
   3 & $\geq 15$ & $\geq 0.1$ & $\leq 1.0$ & $\geq 400$ & $\geq 0.027$ & $\geq  0$ & $\geq  0$ & $\geq  0$ & $\geq  0$ & 64 & 146 & 109 &101 & 0.467 & 0.568 & 0.526 \\ \hline
   4 & $\geq 15$ & $\geq 0.1$ & $\leq 1.0$ & $\geq 500$ & $\geq 0.025$ & $\geq  0$ & $\geq  0$ & $\geq  0$ & $\geq  0$ & 65 & 170 & 111 &124 & 0.435 & 0.543 & 0.497 \\
   5 & $\geq 15$ & $\geq 0.1$ & $\leq 1.0$ & $\geq 500$ & $\geq 0.026$ & $\geq  0$ & $\geq  0$ & $\geq  0$ & $\geq  0$ & 64 & 149 & 107 &106 & 0.462 & 0.553 & 0.515 \\
   6 & $\geq 15$ & $\geq 0.1$ & $\leq 1.0$ & $\geq 500$ & $\geq 0.027$ & $\geq  0$ & $\geq  0$ & $\geq  0$ & $\geq  0$ & 61 & 128 & 101 & 88 & 0.482 & 0.556 & 0.526 \\ \hline
   
   7 & $\geq 15$ & $\geq 0.1$ & $\leq 0.8$ & $\geq 400$ & $\geq 0.025$ & $\geq  0$ & $\geq  0$ & $\geq  0$ & $\geq  0$ & 68 & 161 & 113 &116 & 0.464 & 0.561 & 0.520 \\
   8 & $\geq 15$ & $\geq 0.1$ & $\leq 0.8$ & $\geq 400$ & $\geq 0.026$ & $\geq  0$ & $\geq  0$ & $\geq  0$ & $\geq  0$ & 67 & 144 & 109 &102 & 0.487 & 0.566 & 0.533 \\
   9 & $\geq 15$ & $\geq 0.1$ & $\leq 0.8$ & $\geq 400$ & $\geq 0.027$ & $\geq  0$ & $\geq  0$ & $\geq  0$ & $\geq  0$ & 64 & 124 & 103 & 85 & 0.508 & 0.569 & 0.544 \\ \hline
   10& $\geq 15$ & $\geq 0.1$ & $\leq 0.8$ & $\geq 500$ & $\geq 0.025$ & $\geq  0$ & $\geq  0$ & $\geq  0$ & $\geq  0$ & 65 & 144 & 106 &103 & 0.476 & 0.554 & 0.521 \\
   11& $\geq 15$ & $\geq 0.1$ & $\leq 0.8$ & $\geq 500$ & $\geq 0.026$ & $\geq  0$ & $\geq  0$ & $\geq  0$ & $\geq  0$ & 64 & 127 & 102 & 89 & 0.502 & 0.559 & 0.535 \\
   12& $\geq 15$ & $\geq 0.1$ & $\leq 0.8$ & $\geq 500$ & $\geq 0.027$ & $\geq  0$ & $\geq  0$ & $\geq  0$ & $\geq  0$ & 61 & 107 &  96 & 72 & 0.526 & 0.561 & 0.547 \\ \hline
   
   13& $\geq 17$ & $\geq 0.1$ & $\leq 1.0$ & $\geq 400$ & $\geq 0.025$ & $\geq  0$ & $\geq  0$ & $\geq  0$ & $\geq  0$ & 63 & 139 & 106 &96 & 0.474 & 0.564 & 0.526 \\
   14& $\geq 17$ & $\geq 0.1$ & $\leq 1.0$ & $\geq 400$ & $\geq 0.026$ & $\geq  0$ & $\geq  0$ & $\geq  0$ & $\geq  0$ & 62 & 128 & 105 &85 & 0.488 & 0.577 & 0.540 \\
   15& $\geq 17$ & $\geq 0.1$ & $\leq 1.0$ & $\geq 400$ & $\geq 0.027$ & $\geq  0$ & $\geq  0$ & $\geq  0$ & $\geq  0$ & 60 & 114 & 100 &74 & 0.504 & 0.575 & 0.546 \\ \hline
   16& $\geq 17$ & $\geq 0.1$ & $\leq 1.0$ & $\geq 500$ & $\geq 0.025$ & $\geq  0$ & $\geq  0$ & $\geq  0$ & $\geq  0$ & 61 & 130 & 102 &89 & 0.478 & 0.559 & 0.526 \\
   17& $\geq 17$ & $\geq 0.1$ & $\leq 1.0$ & $\geq 500$ & $\geq 0.026$ & $\geq  0$ & $\geq  0$ & $\geq  0$ & $\geq  0$ & 60 & 119 & 101 &78 & 0.494 & 0.572 & 0.540 \\
   18& $\geq 17$ & $\geq 0.1$ & $\leq 1.0$ & $\geq 500$ & $\geq 0.027$ & $\geq  0$ & $\geq  0$ & $\geq  0$ & $\geq  0$ & 58 & 105 &  96 &67 & 0.511 & 0.570 & 0.546 \\ \hline
   
   19& $\geq 17$ & $\geq 0.1$ & $\leq 0.8$ & $\geq 400$ & $\geq 0.025$ & $\geq  0$ & $\geq  0$ & $\geq  0$ & $\geq  0$ & 63 & 120 & 101 &82 & 0.510 & 0.566 & 0.543 \\
   20& $\geq 17$ & $\geq 0.1$ & $\leq 0.8$ & $\geq 400$ & $\geq 0.026$ & $\geq  0$ & $\geq  0$ & $\geq  0$ & $\geq  0$ & 62 & 112 & 100 &74 & 0.521 & 0.575 & 0.553 \\
   21& $\geq 17$ & $\geq 0.1$ & $\leq 0.8$ & $\geq 400$ & $\geq 0.027$ & $\geq  0$ & $\geq  0$ & $\geq  0$ & $\geq  0$ & 60 &  98 & 95 & 63 & 0.541 & 0.572 & 0.560 \\ \hline
   22& $\geq 17$ & $\geq 0.1$ & $\leq 0.8$ & $\geq 500$ & $\geq 0.025$ & $\geq  0$ & $\geq  0$ & $\geq  0$ & $\geq  0$ & 61 & 111 & 97 & 75 & 0.517 & 0.561 & 0.543 \\
   23& $\geq 17$ & $\geq 0.1$ & $\leq 0.8$ & $\geq 500$ & $\geq 0.026$ & $\geq  0$ & $\geq  0$ & $\geq  0$ & $\geq  0$ & 60 & 103 & 96 & 67 & 0.529 & 0.570 & 0.553 \\
   24& $\geq 17$ & $\geq 0.1$ & $\leq 0.8$ & $\geq 500$ & $\geq 0.027$ & $\geq  0$ & $\geq  0$ & $\geq  0$ & $\geq  0$ & 58 &  89 & 91 & 56 & 0.550 & 0.567 & 0.560 \\ \hline

   25& $\geq 15$ & $\geq 0.1$ & $\leq 0.8$ & $\geq 400$ & $\geq 0.025$ & $\geq 7$ & $\geq 50$ & $\geq 1$ & $\geq  15$ & 67 & 149 & 110 & 106 & 0.479 & 0.564 & 0.528 \\
   26& $\geq 15$ & $\geq 0.1$ & $\leq 0.8$ & $\geq 400$ & $\geq 0.025$ & $\geq 7$ & $\geq 70$ & $\geq 1$ & $\geq  15$ & 66 & 145 & 109 & 101 & 0.480 & 0.568 & 0.531 \\ 
   27& $\geq 15$ & $\geq 0.1$ & $\leq 0.8$ & $\geq 400$ & $\geq 0.025$ & $\geq 1$ & $\geq  15$ & $\geq 2$ & $\geq  26$ & 66 & 128 & 106 &  88 & 0.512 & 0.576 & 0.550 \\
   28& $\geq 15$ & $\geq 0.1$ & $\leq 0.8$ & $\geq 400$ & $\geq 0.025$ & $\geq 1$ & $\geq  15$ & $\geq 2$ & $\geq 30$ & 66 & 117 & 100 &  82 & 0.534 & 0.562 & 0.551 \\ \hline

   29& $\geq 15$ & $\geq 0.1$ & $\leq 0.8$ & $\geq 400$ & $\geq 0.027$ & $\geq 7$ & $\geq 50$ & $\geq 1$ & $\geq  15$ & 63 & 114 & 101 &  76 & 0.523 & 0.575 & 0.554 \\
   30& $\geq 15$ & $\geq 0.1$ & $\leq 0.8$ & $\geq 400$ & $\geq 0.027$ & $\geq 7$ & $\geq 70$ & $\geq 1$ & $\geq  15$ & 62 &  95 & 100 &  73 & 0.561 & 0.576 & 0.570 \\
   31& $\geq 15$ & $\geq 0.1$ & $\leq 0.8$ & $\geq 400$ & $\geq 0.027$ & $\geq 1$ & $\geq  15$ & $\geq 2$ & $\geq 26$ & 62 & 100 &  97 &  65 & 0.549 & 0.577 & 0.566 \\
   32& $\geq 15$ & $\geq 0.1$ & $\leq 0.8$ & $\geq 400$ & $\geq 0.027$ & $\geq 1$ & $\geq  15$ & $\geq 2$ & $\geq 30$ & 61 &  91 &  92 &  60 & 0.565 & 0.564 & 0.565 \\ \hline

   33& $\geq 15$ & $\geq 0.1$ & $\leq 0.8$ & $\geq 400$ & $\geq 0.025$ & $\geq 10$ & $\geq 50$ & $\geq 1$ & $\geq  15$ & 65 & 122 &  99 &  88 & 0.518 & 0.548 & 0.536 \\
   34& $\geq 15$ & $\geq 0.1$ & $\leq 0.8$ & $\geq 400$ & $\geq 0.025$ & $\geq 10$ & $\geq 70$ & $\geq 1$ & $\geq  15$ & 65 & 122 &  99 &  88 & 0.518 & 0.548 & 0.536 \\ 
   35& $\geq 15$ & $\geq 0.1$ & $\leq 0.8$ & $\geq 400$ & $\geq 0.025$ & $\geq 7$ & $\geq  50$ & $\geq 1$ & $\geq 26$ & 65 & 121 & 103 &  83 & 0.520 & 0.572 & 0.551 \\
   36& $\geq 15$ & $\geq 0.1$ & $\leq 0.8$ & $\geq 400$ & $\geq 0.025$ & $\geq 7$ & $\geq  70$ & $\geq 1$ & $\geq 26$ & 64 & 119 & 102 &  81 & 0.518 & 0.571 & 0.550 \\ \hline

   37& $\geq 15$ & $\geq 0.1$ & $\leq 0.8$ & $\geq 400$ & $\geq 0.027$ & $\geq 10$ & $\geq 50$ & $\geq 1$ & $\geq  15$ & 61 &  96 &  93 &  64 & 0.552 & 0.562 & 0.558 \\
   38& $\geq 15$ & $\geq 0.1$ & $\leq 0.8$ & $\geq 400$ & $\geq 0.027$ & $\geq 10$ & $\geq 70$ & $\geq 1$ & $\geq  15$ & 61 &  79 &  93 &  64 & 0.598 & 0.562 & 0.576 \\
   39& $\geq 15$ & $\geq 0.1$ & $\leq 0.8$ & $\geq 400$ & $\geq 0.027$ & $\geq 7$ & $\geq  50$ & $\geq 1$ & $\geq 26$ & 61 &  95 &  95 &  61 & 0.555 & 0.576 & 0.567 \\
   40& $\geq 15$ & $\geq 0.1$ & $\leq 0.8$ & $\geq 400$ & $\geq 0.027$ & $\geq 7$ & $\geq  70$ & $\geq 1$ & $\geq 26$ & 60 &  95 &  94 &  60 & 0.548 & 0.573 & 0.563 \\ \hline

   41& $\geq 15$ & $\geq 0.1$ & $\leq 0.8$ & $\geq 470$ & $\geq 0.027$ & $\geq 10$ & $\geq 50$ & $\geq 1$ & $\geq  15$ & 61 &  92 &  91 &  62 & 0.562 & 0.557 & 0.559 \\
   42& $\geq 15$ & $\geq 0.1$ & $\leq 0.8$ & $\geq 470$ & $\geq 0.027$ & $\geq 10$ & $\geq 70$ & $\geq 1$ & $\geq  15$ & 61 &  76 &  91 &  62 & 0.607 & 0.557 & 0.576 \\
   43& $\geq 15$ & $\geq 0.1$ & $\leq 0.8$ & $\geq 470$ & $\geq 0.027$ & $\geq 7$ & $\geq  50$ & $\geq 1$ & $\geq  26$ & 60 &  91 &  92 &  59 & 0.558 & 0.566 & 0.563 \\
   44& $\geq 15$ & $\geq 0.1$ & $\leq 0.8$ & $\geq 470$ & $\geq 0.027$ & $\geq 7$ & $\geq  70$ & $\geq 1$ & $\geq 26$ & 59 & 91 &  91 &  58 & 0.551 & 0.563 & 0.559 \\ \hline

   45& $\geq$15 & $\geq$ 0.1 & $\leq$0.8 & $\geq$470 & $\geq$0.027 & $\geq$10 & $\geq$100 & $\geq$1 & $\geq 15 $ & 61 & 75 & 91 &  58 & 0.610 & 0.563 & 0.580\\
   \textbf{46}& $\geq$\textbf{15} & $\geq$\textbf{0.1} & $\leq$\textbf{0.8} & $\geq$\textbf{470} & $\geq$\textbf{0.027} & $\geq$\textbf{10} & $\geq$\textbf{120} & $\geq$\textbf{1} & $\geq  $\textbf{15} & \textbf{58} &  \textbf{67} & \textbf{87} &  \textbf{49} & \textbf{0.614} & \textbf{0.561} & \textbf{0.581} \\
   47& $\geq 15$ & $\geq 0.1$ & $\leq 0.8$ & $\geq 470$ & $\geq 0.027$ & $\geq  7$ & $\geq 100$ & $\geq 1$ & $\geq 26$ & 59 &  77 &  86 &  50 & 0.590 & 0.555 & 0.569 \\
   48& $\geq 15$ & $\geq 0.1$ & $\leq 0.8$ & $\geq 470$ & $\geq 0.027$ & $\geq  7$ & $\geq 120$ & $\geq 1$ & $\geq 26$ & 57 &  67 &  82 &  42 & 0.606 & 0.550 & 0.572 \\ \hline

   49& $\geq 15$ & $\geq 0.1$ & $\leq 0.8$ & $\geq 400$ & $\geq 0.027$ & $\geq  5$ & $\geq  50$ & $\geq  2$ & $\geq  26$ & 62 & 98 & 96 & 64 & 0.554 & 0.575 & 0.566 \\ \hline \hline

   50& $\geq 15$ & $\geq 0.1$ & $\leq 0$ & $\geq 0$ & $\geq 0$ & $\geq 0$ & $\geq 0$ & $\geq 0$ & $\geq  0$ & 76 &  284 &  147 &  213 & 0.358 & 0.551 & 0.466 \\
   51& $\geq 0$ & $\geq 0.1$ & $\leq 0$ & $\geq 470$ & $\geq 0$ & $\geq 0$ & $\geq 0$ & $\geq 0$ & $\geq  0$ & 79 &  364 &  151 &  292 & 0.312 & 0.489 & 0.409 \\
   52& $\geq 0$ & $\geq 0.1$ & $\leq 0$ & $\geq 0$ & $\geq 0.027$ & $\geq 0$ & $\geq 0$ & $\geq 0$ & $\geq  0$ & 70 &  227 &  127 &  372 & 0.388 & 0.377 & 0.381 \\
   53& $\geq 0$ & $\geq 0.1$ & $\leq 0$ & $\geq 0$ & $\geq 0$ & $\geq 10$ & $\geq 0$ & $\geq 0$ & $\geq  0$ & 75 &  267 &  133 &  209 & 0.369 & 0.516 & 0.451 \\
   54& $\geq 0$ & $\geq 0.1$ & $\leq 0$ & $\geq 0$ & $\geq 0$ & $\geq 0$ & $\geq 120$ & $\geq 0$ & $\geq  0$ & 67 &  199 &  130 &  186 & 0.406 & 0.531 & 0.480 \\ \hline
    \end{tabular}
    \end{center}
\textbf{Notes.} We distinguish between lens candidates classified as clusters during visual inspection (see Sect.~\ref{sec:environment:inspection}) and based on galaxy catalogs \citep{oguri14, oguri18, wen18, wen21}. Without photo-$z$ selection (see No. 0), we visually identified 84 lens candidates to be in an overdense environment, while 174 lens candidates are located closer than $100\arcsec$ from a known galaxy cluster. We highlight No.~46 in bold, which has the selection criteria cuts that are the best according to the F1$_\text{tot}$ statistic.
    \label{tab:photozstat}
\end{table*}

In this appendix, we provide in Table~\ref{tab:photozstat} the details of our lens cluster selection regarding their environment described in Sect.~\ref{sec:environment}. We provide further details of our lens cluster selection regarding their environment. In detail, Table~\ref{tab:photozstat_subtr} gives, in analogy to Table~\ref{tab:photozstat}, the performance of photo-$z$ cuts on the subtracted photo-$z$ histograms, and shows that No. 43 lists the best criteria limits according to the F1$_\text{tot}$ value. We note that this selection is outperformed by No. 46 presented in Sect.~\ref{sec:environment:phototz} using the original histograms, instead of the subtracted ones. 

\begin{figure*}
\centering
  \includegraphics[trim=0 0 0 0, clip, width=\textwidth]{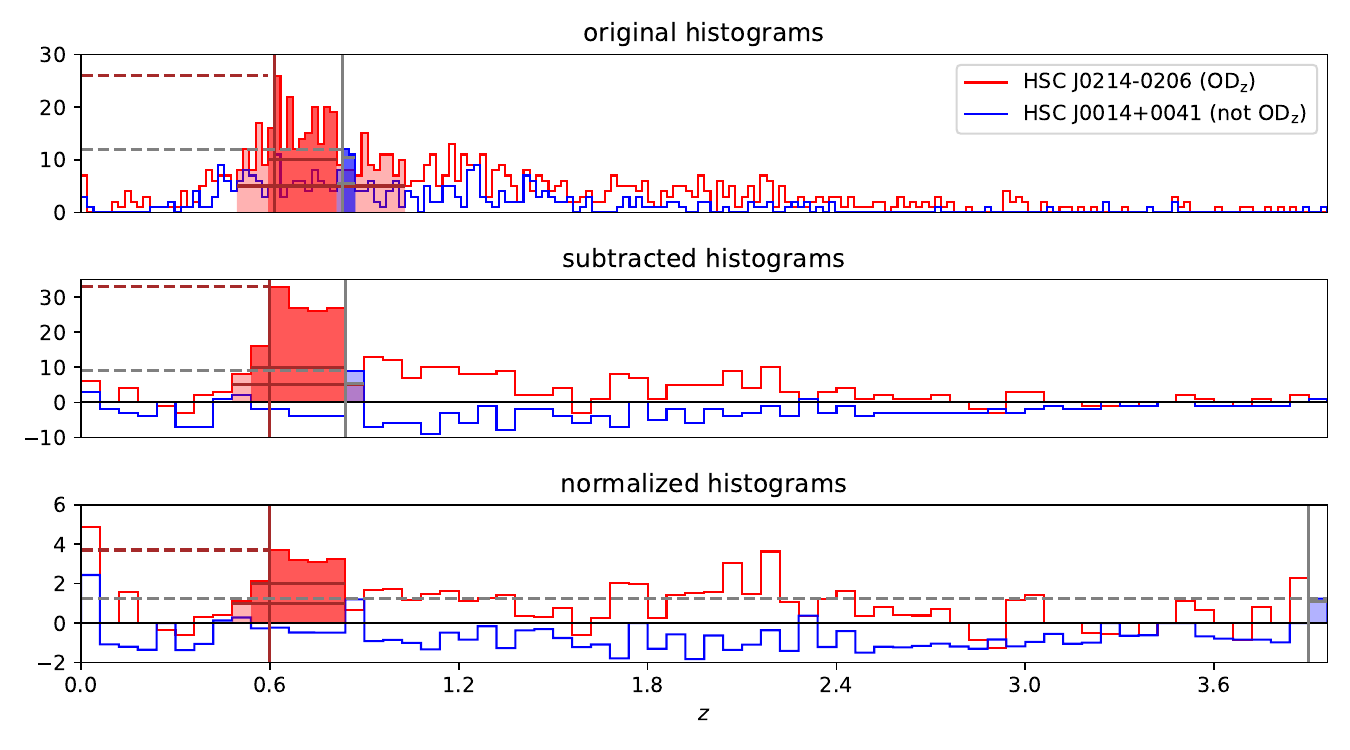}
  \caption{Redshift histograms of two lens candidates as example, one visually identified to be in an overdense environment (HSC J0214-0206, red) and one not (HSC J0014+0041, blue). We show the original histograms (top panel), the histograms where the median distribution of 10 000 random sightlines are subtracted from the original (middle panel), and the normalized histograms (bottom panel) where the subtracted histograms are normalized by the standard deviation of the 10 000 sightlines. Our defined criteria, introduced in Sect.~\ref{sec:environment:phototz}, are indicated with brown (OD$_\text{vis}$) and gray (not OD$_\text{vis}$) lines. In detail, the vertical solid line indicates $z_\text{low}$, the lower bound of the tallest redshift bin of the histogram. The height of this bin, denoted as $N_\text{max}$, is marked by the dashed horizontal line. The lengths of the horizontal solid lines, which are the number of the shaded bins, represent $N_\text{peak5}$ and $N_\text{peak10}$, while the sum of objects in all the shaded bins corresponds to the value of $A_\text{peak5}$ and $A_\text{peak10}$. See Appendix \ref{sec:appendixB} for further details.}
  \label{fig:hist_examples}
\end{figure*}

In Fig.~\ref{fig:hist_examples} we show the photo-$z$ histograms of two lens candidates as example, one visually identified to be in an overdense environment (HSC J0214-0206, red) and one not (HSC J0014+0041, blue). In detail, we show the original histograms (top panel), the subtracted histograms (middle panel), and the normalized histograms (bottom panel). The eight selection criteria, introduced in Sect.~\ref{sec:environment:phototz}, are marked for visualization. For instance, HSC J0214-0206 has with the original histogram (top panel) a value for N$_\text{peak}$ of 26 (horizontal dashed brown line) at $z_\text{low}=0.62$ (vertical solid brown line). The sum of the red histogram results in N$_\text{tot}=810$, and consequently N$_\text{frac}=26/810\sim 0.032$. In addition, this histogram shows N$_\text{peak5}=26$ and N$_\text{peak10}=11$ neighbouring redshift bins around $z_\text{low}$ that exceeds 5 and 10 (horizontal solid brown line). The sums of objects in these redshift bins are 336 and 189, respectively, denoted as A$_\text{peak5}$ and A$_\text{peak10}$, that are shaded in light and solid red. Following the same style, we also indicate the values of the criteria for the other histograms. 

\begin{figure*}
\centering
  \includegraphics[trim=0 0 0 0, clip, width=\textwidth]{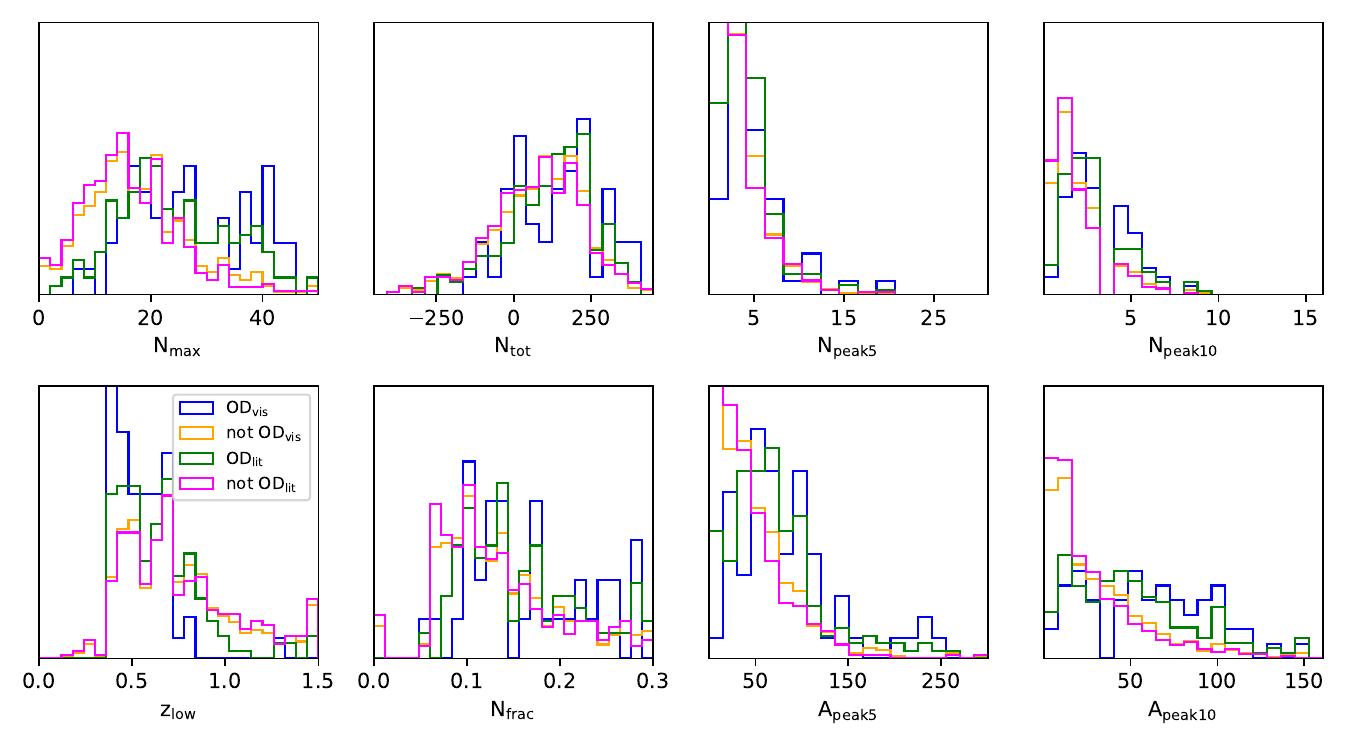}
  \caption{Normalized histograms of our eight introduced selection criteria using the subtracted photo-$z$ distributions. We distinguish between lens candidates visually identified to be in an overdensity (blue) or through the galaxy cluster catalogs (green), compared to those not in an overdensity (orange and magenta, respectively).}
  \label{fig:photozcomparison_subtrhist}
\end{figure*}

\begin{figure*}
\centering
  \includegraphics[trim=0 0 0 0, clip, width=\textwidth]{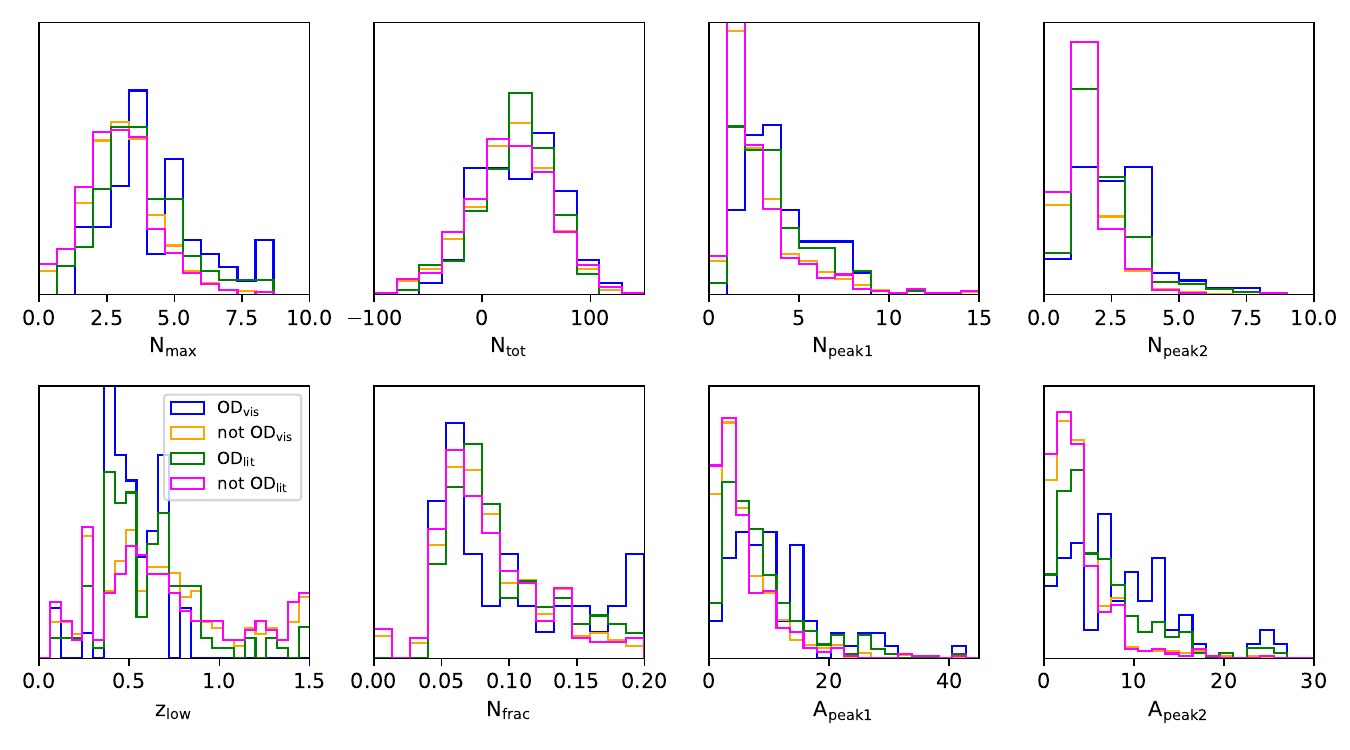}
  \caption{Normalized histograms of our eight introduced selection criteria using the normalized photo-$z$ distributions (which are the subtracted histograms in Fig.~\ref{fig:hist_examples}, normalized by the standard deviation of the 10 000 random sightlines' histograms). We distinguish between lens candidates visually identified to be in an overdensity (blue) or through the galaxy cluster catalogs (green), compared to those not in an overdensity (orange and magenta, respectively).}
  \label{fig:photozcomparison_normhist}
\end{figure*}

\begin{figure*}
\centering
  \includegraphics[trim=0 0 0 0, clip, width=\textwidth]{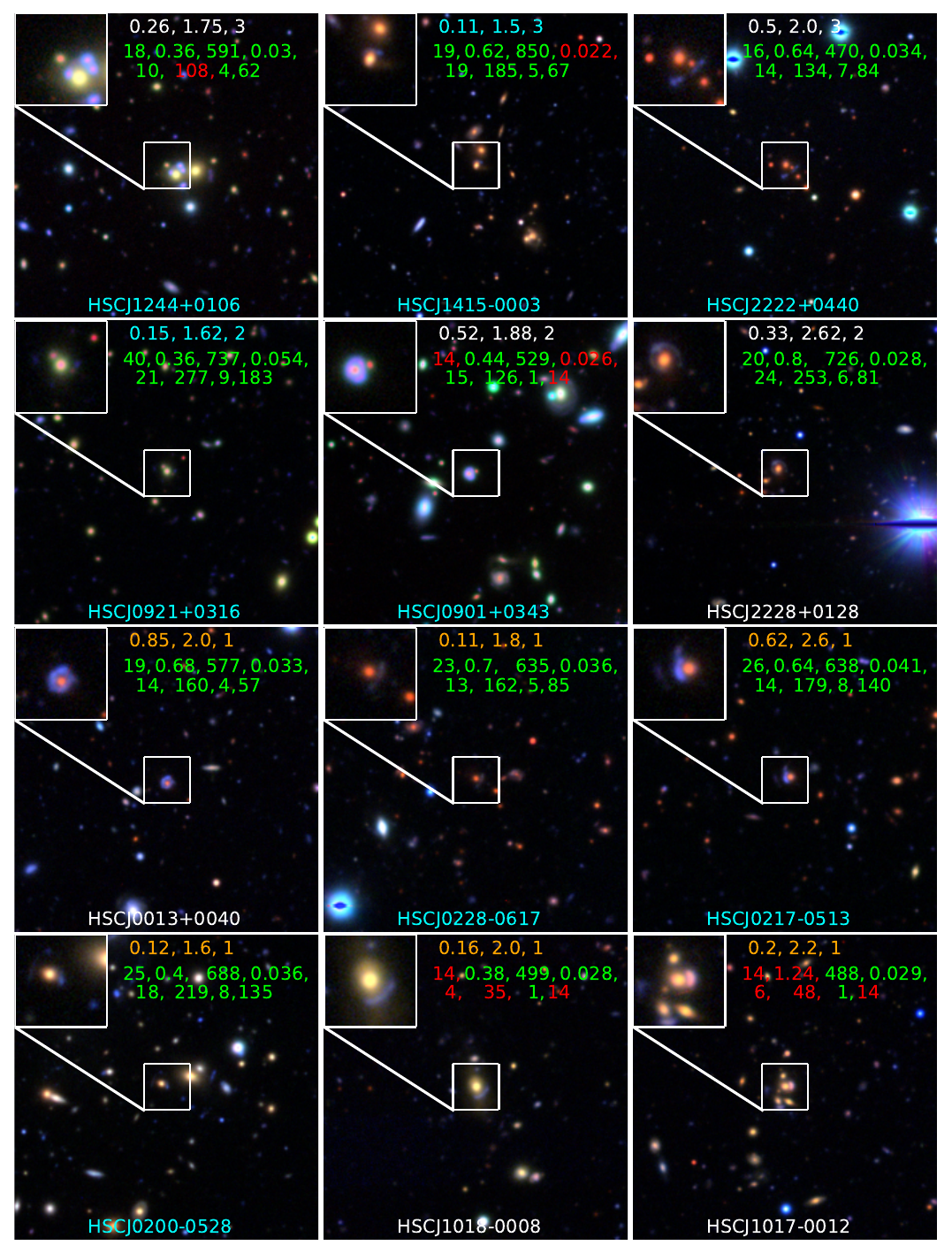}
  \caption{New grade A and B group- or cluster lens candidates based on visual inspection. Continuation of Fig.~\ref{fig:clusterexamples}.}
  \label{fig:clusterexamples_appendix}
\end{figure*}

Furthermore, Fig.~\ref{fig:photozcomparison_subtrhist} visualizes the introduced criteria using the subtracted histograms, while  Fig.~\ref{fig:photozcomparison_normhist} shows them extracted from the normalized histograms. Finally, we show in Fig.~\ref{fig:clusterexamples_appendix} $80\arcsec \times 80\arcsec$ image stamps of the remaining grade A and B lens candidates located in an overdense environment based on our visual inspection. This is the continuation of Fig.~\ref{fig:clusterexamples}.

\begin{table*}[p]
    \caption{Comparison of true-positive (TP) and false-positive (FP) rates for different selection criteria using the photometric redshift distributions around the 546 grade A or B lens candidates after subtracting a median distribution from 10 000 random sky positions.}
    \begin{center}
    \begin{tabular}{c|ccccccccc|cc|cc|ccc}
   No. &\multicolumn{9}{c|}{Criteria} & \multicolumn{2}{c|}{visual insp.} & \multicolumn{2}{c|}{Literature} & \multicolumn{3}{c}{F1 statistics}\\ \hline \noalign{\smallskip}
   & $N_\text{max}$ & $z_\text{low}$ & $z_\text{low}$& $N_\text{tot}$ & $N_\text{frac}$ & $N_\text{peak5}$ & $A_\text{peak5}$ & $N_\text{peak10}$ & $A_\text{peak10}$ & TP & FP & TP & FP & F1$_\text{vis}$ & F1$_\text{lit}$ & F1$_\text{tot}$ \\ 
\hline \hline \noalign{\smallskip}
   1 & $\geq 11$ & $\geq 0.1$ & $\leq 1.0$ & $\geq  0$ & $\geq 0.07$ & $\geq  0 $& $\geq  0$ & $\geq  0$ & $\geq  0$ & 69 & 240 & 134 & 175 & 0.370 & 0.555 & 0.474 \\
   2 & $\geq 11$ & $\geq 0.1$ & $\leq 1.0$ & $\geq  0$ & $\geq 0.10$ & $\geq  0$ & $\geq  0$ & $\geq  0$ & $\geq  0$ & 66 & 194 & 118 & 142 & 0.407 & 0.544 & 0.485 \\
   3 & $\geq 11$ & $\geq 0.1$ & $\leq 1.0$ & $\geq  0$ & $\geq 0.13$ & $\geq  0 $& $\geq  0$ & $\geq  0$ & $\geq  0$ & 53 & 147 &  98 & 101 & 0.402 & 0.525 & 0.474 \\ \hline
   4 & $\geq 11$ & $\geq 0.1$ & $\leq 1.0$ & $\geq 50$ & $\geq 0.07$ & $\geq  0 $& $\geq  0$ & $\geq  0$ & $\geq  0$ & 60 & 222 & 116 & 155 & 0.347 & 0.521 & 0.445 \\
   5 & $\geq 11$ & $\geq 0.1$ & $\leq 1.0$ & $\geq 50$ & $\geq 0.10$ & $\geq  0 $& $\geq  0$ & $\geq  0$ & $\geq  0$ & 57 & 176 & 100 & 122 & 0.384 & 0.505 & 0.453 \\
   6 & $\geq 11$ & $\geq 0.1$ & $\leq 1.0$ & $\geq 50$ & $\geq 0.13$ & $\geq  0 $& $\geq  0$ & $\geq  0$ & $\geq  0$ & 44 & 129 &  80 &  81 & 0.371 & 0.478 & 0.434 \\ \hline
   
   7 & $\geq 11$ & $\geq 0.1$ & $\leq 0.8$ & $\geq  0$ & $\geq 0.07$ & $\geq  0 $& $\geq  0$ & $\geq  0$ & $\geq  0$ & 69 & 196 & 121 & 144 & 0.419 & 0.551 & 0.495 \\
   8 & $\geq 11$ & $\geq 0.1$ & $\leq 0.8$ & $\geq  0$ & $\geq 0.10$ & $\geq  0 $& $\geq  0$ & $\geq  0$ & $\geq  0$ & 66 & 159 & 106 & 119 & 0.457 & 0.531 & 0.500 \\
   9 & $\geq 11$ & $\geq 0.1$ & $\leq 0.8$ & $\geq  0$ & $\geq 0.13$ & $\geq  0 $& $\geq  0$ & $\geq  0$ & $\geq  0$ & 53 & 126 &  91 &  87 & 0.436 & 0.517 & 0.484 \\ \hline
   10& $\geq 11$ & $\geq 0.1$ & $\leq 0.8$ & $\geq 50$ & $\geq 0.07$ & $\geq  0 $& $\geq  0$ & $\geq  0$ & $\geq  0$ & 60 & 181 & 105 & 126 & 0.393 & 0.519 & 0.465 \\
   11& $\geq 11$ & $\geq 0.1$ & $\leq 0.8$ & $\geq 50$ & $\geq 0.10$ & $\geq  0 $& $\geq  0$ & $\geq  0$ & $\geq  0$ & 57 & 144 &  90 & 101 & 0.430 & 0.493 & 0.467 \\
   12& $\geq 11$ & $\geq 0.1$ & $\leq 0.8$ & $\geq 50$ & $\geq 0.13$ & $\geq  0 $& $\geq  0$ & $\geq  0$ & $\geq  0$ & 44 & 111 &  75 &  69 & 0.402 & 0.472 & 0.443 \\ \hline
   
   13& $\geq 16$ & $\geq 0.1$ & $\leq 1.0$ & $\geq  0$ & $\geq 0.07$ & $\geq  0 $& $\geq  0$ & $\geq  0$ & $\geq  0$ & 67 & 187 & 123 & 131 & 0.421 & 0.575 & 0.509 \\
   14& $\geq 16$ & $\geq 0.1$ & $\leq 1.0$ & $\geq  0$ & $\geq 0.10$ & $\geq  0 $& $\geq  0$ & $\geq  0$ & $\geq  0$ & 64 & 148 & 107 & 105 & 0.464 & 0.554 & 0.517 \\
   15& $\geq 16$ & $\geq 0.1$ & $\leq 1.0$ & $\geq  0$ & $\geq 0.13$ & $\geq  0 $& $\geq  0$ & $\geq  0$ & $\geq  0$ & 51 & 108 &  88 &  71 & 0.457 & 0.529 & 0.500 \\ \hline
   16& $\geq 16$ & $\geq 0.1$ & $\leq 1.0$ & $\geq 50$ & $\geq 0.07$ & $\geq  0 $& $\geq  0$ & $\geq  0$ & $\geq  0$ & 60 & 177 & 110 & 122 & 0.399 & 0.542 & 0.481 \\
   17& $\geq 16$ & $\geq 0.1$ & $\leq 1.0$ & $\geq 50$ & $\geq 0.10$ & $\geq  0 $& $\geq  0$ & $\geq  0$ & $\geq  0$ & 57 & 138 &  94 &  96 & 0.440 & 0.516 & 0.485 \\
   18& $\geq 16$ & $\geq 0.1$ & $\leq 1.0$ & $\geq 50$ & $\geq 0.13$ & $\geq  0 $& $\geq  0$ & $\geq  0$ & $\geq  0$ & 44 &  98 &  75 &  62 & 0.427 & 0.482 & 0.460 \\ \hline
   
   19& $\geq 16$ & $\geq 0.1$ & $\leq 0.8$ & $\geq  0$ & $\geq 0.07$ & $\geq  0 $& $\geq  0$ & $\geq  0$ & $\geq  0$ & 67 & 151 & 111 & 107 & 0.475 & 0.566 & 0.528 \\
   20& $\geq 16$ & $\geq 0.1$ & $\leq 0.8$ & $\geq  0$ & $\geq 0.10$ & $\geq  0 $& $\geq  0$ & $\geq  0$ & $\geq  0$ & 64 & 118 &  96 &  86 & 0.520 & 0.539 & 0.532 \\
   21& $\geq 16$ & $\geq 0.1$ & $\leq 0.8$ & $\geq  0$ & $\geq 0.13$ & $\geq  0 $& $\geq  0$ & $\geq  0$ & $\geq  0$ & 51 &  91 &  82 &  60 & 0.495 & 0.519 & 0.510 \\ \hline
   22& $\geq 16$ & $\geq 0.1$ & $\leq 0.8$ & $\geq 50$ & $\geq 0.07$ & $\geq  0 $& $\geq  0$ & $\geq  0$ & $\geq  0$ & 60 & 143 &  99 & 100 & 0.449 & 0.531 & 0.497 \\
   23& $\geq 16$ & $\geq 0.1$ & $\leq 0.8$ & $\geq 50$ & $\geq 0.10$ & $\geq  0 $& $\geq  0$ & $\geq  0$ & $\geq  0$ & 57 & 110 &  84 &  79 & 0.494 & 0.499 & 0.496 \\
   24& $\geq 16$ & $\geq 0.1$ & $\leq 0.8$ & $\geq 50$ & $\geq 0.13$ & $\geq  0 $& $\geq  0$ & $\geq  0$ & $\geq  0$ & 44 &  83 &  70 &  53 & 0.461 & 0.471 & 0.467 \\ \hline

   25& $\geq 11$ & $\geq 0.1$ & $\leq 0.8$ & $\geq  0$ & $\geq 0.07$ & $\geq  3 $& $\geq  20$ & $\geq  0$ & $\geq  0$ & 66 & 151 & 107 & 128 & 0.470 & 0.523 & 0.501 \\
   26& $\geq 11$ & $\geq 0.1$ & $\leq 0.8$ & $\geq  0$ & $\geq 0.07$ & $\geq  3 $& $\geq  30$ & $\geq  0$ & $\geq  0$ & 65 & 146 & 106 & 114 & 0.473 & 0.538 & 0.511 \\
   27& $\geq 11$ & $\geq 0.1$ & $\leq 0.8$ & $\geq 50$ & $\geq 0.07$ & $\geq  0 $& $\geq   0$ & $\geq  2$ & $\geq 24$ & 61 & 121 & 102 &  90 & 0.496 & 0.557 & 0.533 \\
   28& $\geq 11$ & $\geq 0.1$ & $\leq 0.8$ & $\geq 50$ & $\geq 0.07$ & $\geq  0 $& $\geq   0$ & $\geq  2$ & $\geq 30$ & 60 & 114 & 100 &  82 & 0.504 & 0.562 & 0.539 \\ \hline

   29& $\geq 11$ & $\geq 0.1$ & $\leq 0.8$ & $\geq  0$ & $\geq 0.10$ & $\geq  3 $& $\geq  20$ & $\geq  0$ & $\geq  0$ & 63 & 118 &  93 &  88 & 0.514 & 0.524 & 0.520 \\
   30& $\geq 11$ & $\geq 0.1$ & $\leq 0.8$ & $\geq  0$ & $\geq 0.10$ & $\geq  3 $& $\geq  30$ & $\geq  0$ & $\geq  0$ & 62 & 113 &  92 &  83 & 0.519 & 0.527 & 0.524 \\
   31& $\geq 11$ & $\geq 0.1$ & $\leq 0.8$ & $\geq 50$ & $\geq 0.10$ & $\geq  0 $& $\geq   0$ & $\geq  2$ & $\geq 24$ & 58 &  88 &  89 &  67 & 0.552 & 0.539 & 0.544 \\
   32& $\geq 11$ & $\geq 0.1$ & $\leq 0.8$ & $\geq 50$ & $\geq 0.10$ & $\geq  0 $& $\geq   0$ & $\geq  2$ & $\geq 30$ & 57 &  83 &  87 &  61 & 0.559 & 0.540 & 0.548 \\ \hline

   33& $\geq 11$ & $\geq 0.1$ & $\leq 0.8$ & $\geq  0$ & $\geq 0.07$ & $\geq  3 $& $\geq  35$ & $\geq  0$ & $\geq  0$ & 65 & 133 & 104 & 97 & 0.496 & 0.555 & 0.531 \\
   34& $\geq 11$ & $\geq 0.1$ & $\leq 0.8$ & $\geq  0$ & $\geq 0.07$ & $\geq  3 $& $\geq  40$ & $\geq  0$ & $\geq  0$ & 64 & 123 & 102 & 87 & 0.510 & 0.562 & 0.541 \\
   35& $\geq 11$ & $\geq 0.1$ & $\leq 0.8$ & $\geq 50$ & $\geq 0.07$ & $\geq  2 $& $\geq  20$ & $\geq  0$ & $\geq 24$ & 58 & 128 &  97 & 89 & 0.464 & 0.539 & 0.508 \\
   36& $\geq 11$ & $\geq 0.1$ & $\leq 0.8$ & $\geq 50$ & $\geq 0.07$ & $\geq  2 $& $\geq  20$ & $\geq  0$ & $\geq 30$ & 58 & 115 &  93 & 80 & 0.489 & 0.536 & 0.517 \\ \hline

   37& $\geq 11$ & $\geq 0.1$ & $\leq 0.8$ & $\geq  0$ & $\geq 0.10$ & $\geq  3 $& $\geq  35$ & $\geq  0$ & $\geq  0$ & 62 & 100 &  90 &  75 & 0.549 & 0.531 & 0.538 \\
   38& $\geq 11$ & $\geq 0.1$ & $\leq 0.8$ & $\geq  0$ & $\geq 0.10$ & $\geq  3 $& $\geq  40$ & $\geq  0$ & $\geq  0$ & 61 &  91 &  88 &  66 & 0.565 & 0.537 & 0.548 \\
   39& $\geq 11$ & $\geq 0.1$ & $\leq 0.8$ & $\geq 50$ & $\geq 0.10$ & $\geq  2 $& $\geq  20$ & $\geq  0$ & $\geq 24$ & 55 &  94 &  84 &  65 & 0.516 & 0.520 & 0.519 \\
   40& $\geq 11$ & $\geq 0.1$ & $\leq 0.8$ & $\geq 50$ & $\geq 0.10$ & $\geq  2 $& $\geq  20$ & $\geq  0$ & $\geq 30$ & 55 &  84 &  80 &  59 & 0.542 & 0.511 & 0.523 \\ \hline

   41& $\geq 15$ & $\geq 0.1$ & $\leq 0.8$ & $\geq 50$ & $\geq 0.10$ & $\geq  0 $& $\geq   0$ & $\geq  2$ & $\geq 24$ & 57 &  83 &  87 &  62 & 0.559 & 0.539 & 0.546 \\
   42& $\geq 15$ & $\geq 0.1$ & $\leq 0.8$ & $\geq 50$ & $\geq 0.10$ & $\geq  0 $& $\geq   0$ & $\geq  2$ & $\geq 30$ & 57 &  80 &  86 &  59 & 0.567 & 0.539 & 0.550 \\ \hline
   
   43& $\geq 15$ & $\geq 0.1$ & $\leq 0.8$ & $\geq 50$ & $\geq 0.10$ & $\geq  0 $& $\geq   35$ & $\geq  2$ & $\geq 24$ & 57 & 80 &  86 &  58 & 0.567 & 0.541 & 0.551 \\
   44& $\geq 15$ & $\geq 0.1$ & $\leq 0.8$ & $\geq 50$ & $\geq 0.10$ & $\geq  0 $& $\geq   45$ & $\geq  2$ & $\geq 24$ & 56 & 74 &  82 &  52 & 0.577 & 0.532 & 0.550 \\
   45& $\geq 15$ & $\geq 0.1$ & $\leq 0.8$ & $\geq 50$ & $\geq 0.10$ & $\geq  0 $& $\geq   35$ & $\geq  2$ & $\geq 30$ & 57 & 78 &  85 &  57 & 0.573 & 0.538 & 0.551 \\
   46& $\geq 15$ & $\geq 0.1$ & $\leq 0.8$ & $\geq 50$ & $\geq 0.10$ & $\geq  0 $& $\geq   45$ & $\geq  2$ & $\geq 30$ & 56 & 73 &  81 &  52 & 0.580 & 0.528 & 0.548 \\ \hline
   
    \end{tabular}
    \end{center}
\textbf{Notes.} We distinguish between lens candidates classified as clusters during visual inspection (see Sect.~\ref{sec:environment:inspection}) and based on galaxy catalogs \citep{oguri14, oguri18, wen18, wen21}. Without photo-$z$ selection (see No. 0 in Table~\ref{tab:photozstat}), we visually identified 84 lens candidates to be in an overdense environment, while 174 lens candidates are located closer than $100\arcsec$ from a known galaxy cluster.
    \label{tab:photozstat_subtr}
\end{table*}

\begin{figure*}
\centering
  \includegraphics[trim=0 0 0 0, clip, width=\textwidth]{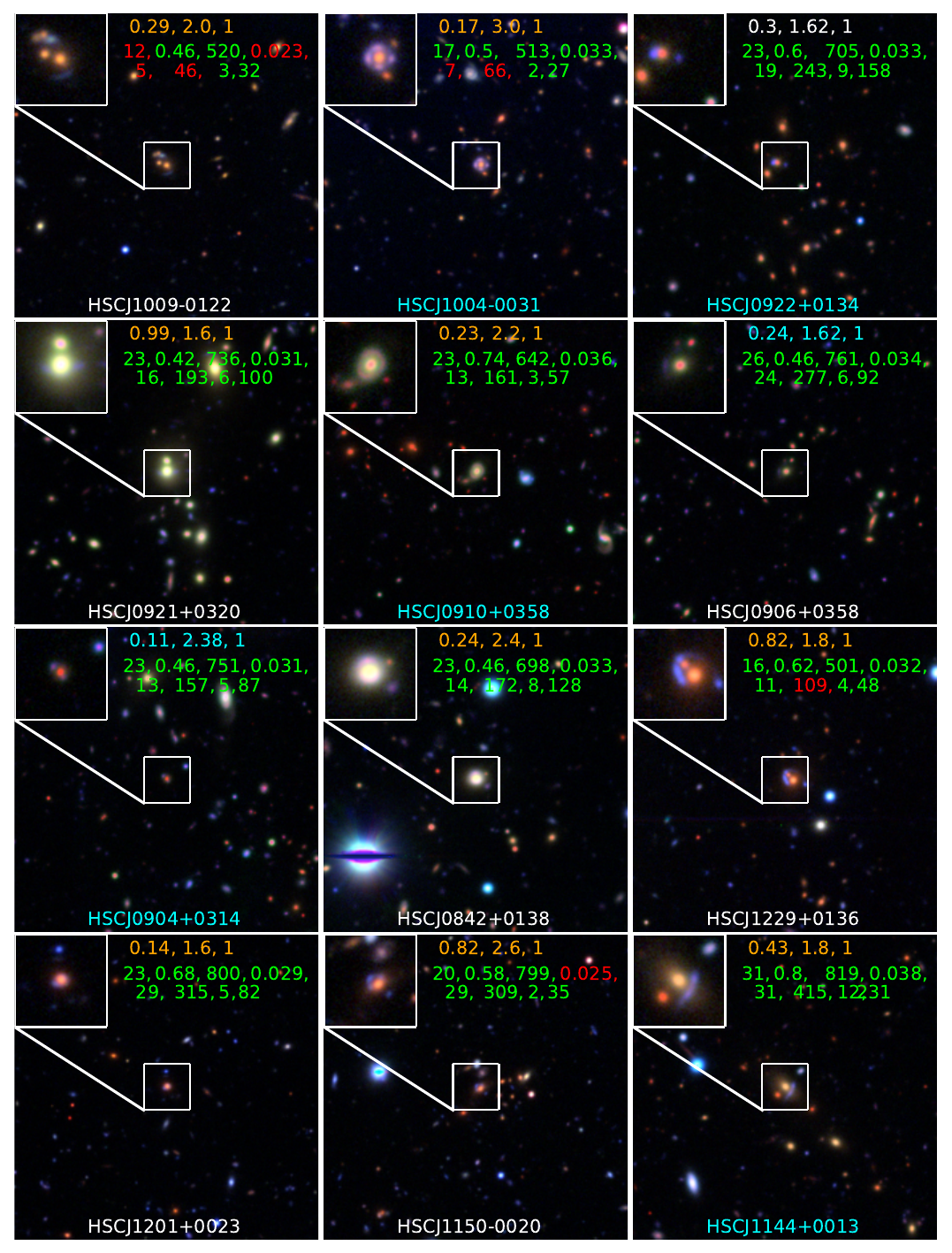}
  \caption*{Fig.~\ref{fig:clusterexamples_appendix} continued: New grade A and B group- or cluster lens candidates based on visual inspection.}
\end{figure*}

\begin{figure*}
\centering
  \includegraphics[trim=0 0 0 0, clip, width=\textwidth]{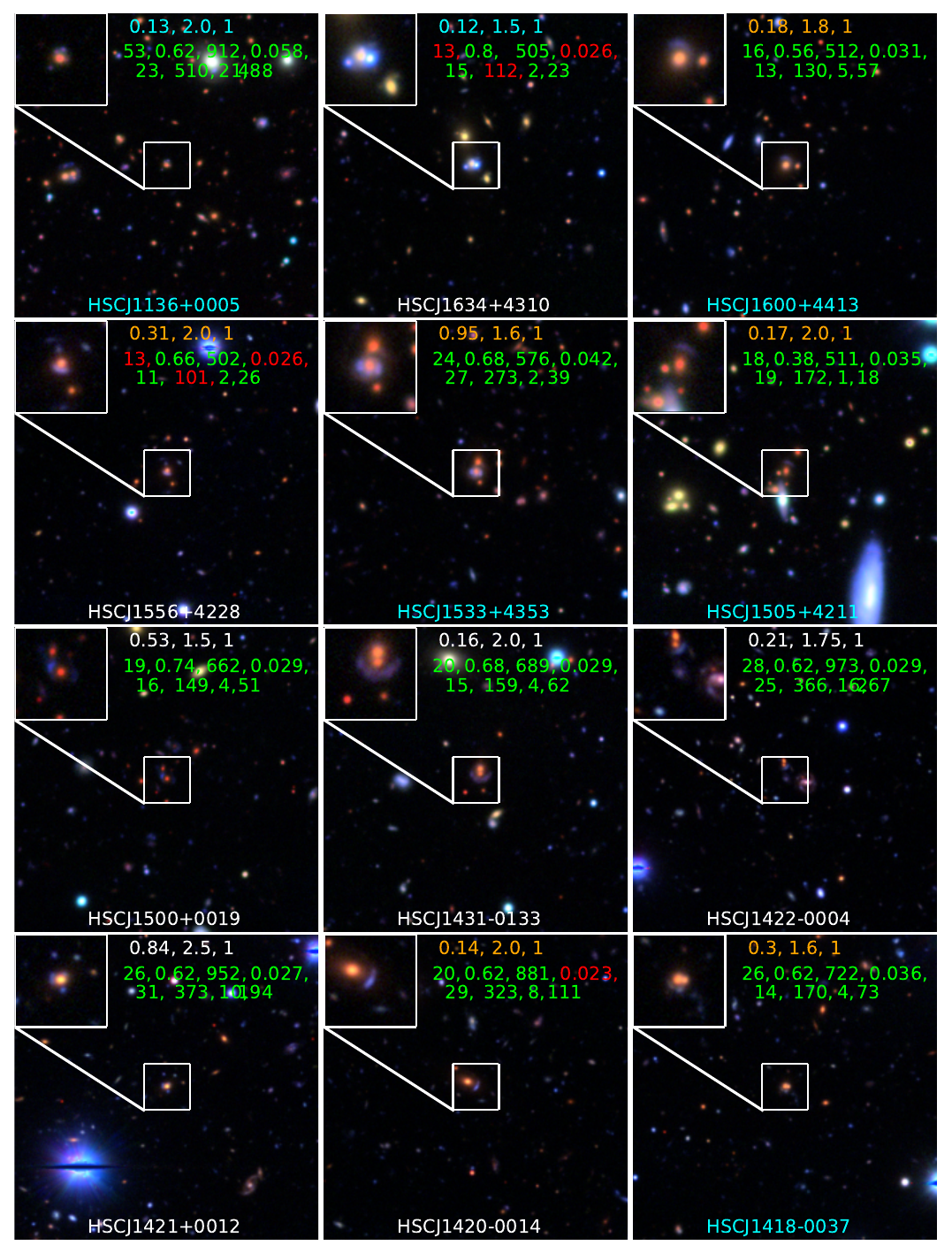}
  \caption*{Fig.~\ref{fig:clusterexamples_appendix} continued: New grade A and B group- or cluster lens candidates based on visual inspection.}
\end{figure*}

\begin{figure*}
\centering
  \includegraphics[trim=0 0 0 0, clip, width=\textwidth]{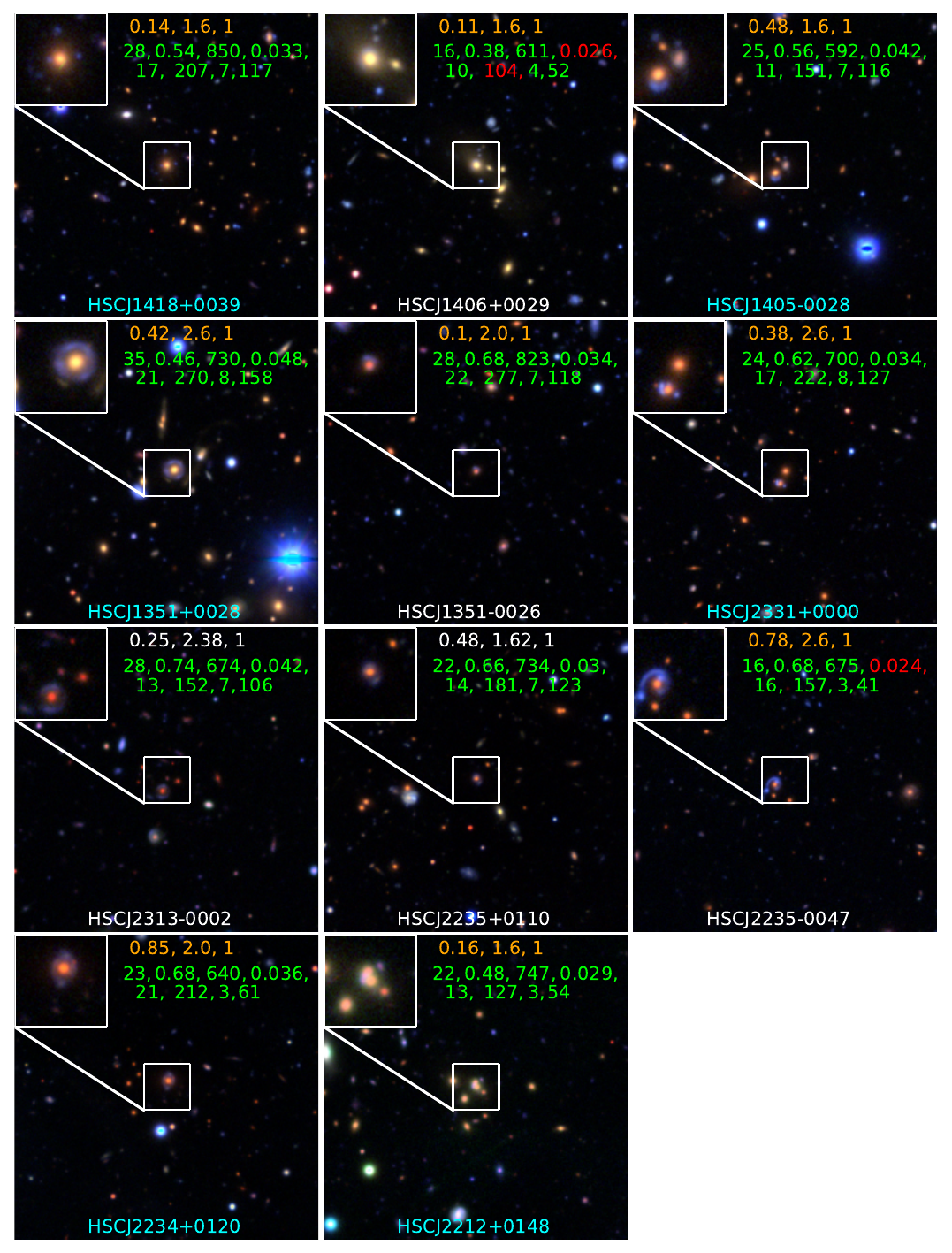}
  \caption*{Fig.~\ref{fig:clusterexamples_appendix} continued: New grade A and B group- or cluster lens candidates based on visual inspection.}
\end{figure*}

\end{document}